\documentclass[12pt]{article}
\usepackage{amsmath}
\usepackage{graphicx,psfrag,epsf}
\usepackage{enumerate}
\usepackage{natbib}
\usepackage{url} 

\newcommand{\blind}{0}

\usepackage{soul}
\usepackage{xcolor}     

\usepackage{calc}
\newsavebox\CBox
\newcommand\hcancel[2][0.5pt]{%
  \ifmmode\sbox\CBox{$#2$}\else\sbox\CBox{#2}\fi%
  \makebox[0pt][l]{\usebox\CBox}%
  \rule[0.5\ht\CBox-#1/2]{\wd\CBox}{#1}}

\begin{document}

\bibliographystyle{agsm}

\def\spacingset#1{\renewcommand{\baselinestretch}%
{#1}\small\normalsize} \spacingset{1}


\if0\blind
{
  \title{\bf The world of research has gone berserk: modeling the consequences of requiring ``greater statistical stringency'' for scientific publication}
  \author{Harlan Campbell
    and 
    Paul Gustafson \thanks{
    The authors gratefully acknowledge Prof. Will Welch and Prof. John Petkau for their valuable suggestions and advice.} \\
    Department of Statistics, University of British Columbia, Vancouver, Canada}
  \maketitle
} \fi

\if1\blind
{
  \bigskip
  \bigskip
  \bigskip
  \begin{center}
    {\LARGE\bf Title}
\end{center}
  \medskip
} \fi

\bigskip
\begin{abstract}
In response to growing concern about the reliability and reproducibility of published science, researchers have proposed adopting measures of `greater statistical stringency', including suggestions to require larger sample sizes and to lower the highly criticized `$p<0.05$' significance threshold.  While pros and cons are vigorously debated, there has been little to no modeling of how adopting these measures might affect what type of science is published.  In this paper, we develop a novel optimality model that, given current incentives to publish, predicts a researcher's most rational use of resources in terms of the number of studies to undertake, the statistical power to devote to each study, and the desirable pre-study odds to pursue.  We then develop a methodology that allows one to estimate the reliability of published research by considering a distribution of preferred research strategies.  Using this approach, we investigate the merits of adopting measures of `greater statistical stringency' with the goal of informing the ongoing debate.
\end{abstract}

\noindent%
\vfill

\newpage
\spacingset{1.45} 

\section{Introduction}
\label{sec:intro}

\small{\begin{quote}{\footnotesize{It is to be remarked that the theory here given rests on the supposition that the object of the investigation is the ascertainment of truth. When an investigation is made for the purpose of attaining personal distinction, the economics of the problem are entirely different. But that seems to be well enough understood by those engaged in that sort of investigation.}}\end{quote}
\hfill{\footnotesize{\emph{Note on the Theory of the Economy of Research, $\quad$ }}}\\
\vspace{-0.75cm}
\hfill {\footnotesize{ Charles Sanders Peirce, 1879}}}
 \vspace{1cm}

 In a highly cited essay, \cite{ioannidis2005most} uses Bayes theorem to claim that more than half of published research findings are false.  While not all agree with the extent of this conclusion (e.g. \cite{goodman2007assessing, leek2017most}), recent large-scale efforts to reproduce published results in a number of different fields (economics, \cite{camerer2016evaluating}; psychology, \cite{open2015estimating}; oncology, \cite{begley2012drug}), have also raised concerns about the reliability and reproducibility of published science.  Unreliable research not only reduces the credibility of science, but is also very costly \citep{freedman2015economics} and as such, addressing the underlying issues is of ``vital importance'' \citep{spiegelhalter2017trust}.  Many researchers have recently proposed adopting measures of ``greater statistical stringency'', including suggestions to require larger sample sizes and to lower the highly criticized  ``$p<0.05$'' significance threshold.  In statistical terms, this represents selecting lower levels for acceptable type I and type II error.

Consider the debate about lowering the significance threshold in response to the work of \cite{johnson2013revised}, who, based on the correspondence between uniformly most powerful Bayesian tests and classical significance tests, recommends lowering significance thresholds by a factor of 10 (e.g. from $p<0.05$ to $p<0.005$).   \cite{gaudart2014reproducibility}, voicing a common objection, contend that such a reduction in the allowable type I error will result in inevitable increases to the type II error.  While larger sample sizes could compensate, this can be costly: ``increasing the size of clinical trials will reduce their feasibility and increase their duration'' \citep{gaudart2014reproducibility}.  In \cite{johnson2014reply}'s view, this may not necessarily be such a bad thing, pointing to the excess of false positives and the idea that (in the context of clinical trials) ``too many ineffective drugs are subjected to phase III testing [...] wast[ing] enormous human and financial resources''.

More recently, a highly publicized call by over seven dozen authors  to ``redefine statistical significance'' has made a similar suggestion: lower the threshold of what is considered ``significant'' from $p\le0.05$ to $p\le0.005$ \citep{benjamin2017redefine}. This has prompted a familiar response.  \cite{amrhein2017earth} review the arguments for and against more stringent thresholds for significance and conclude that: ``[v]ery possibly, more stringent thresholds would lead to even more results being left unpublished, enhancing publication bias. [...] [W]hile aiming at making our published claims more reliable, requesting more stringent fixed thresholds would achieve quite the opposite.''

There is also substantial disagreement about suggestions to require larger sample sizes.  In some fields, showing that a study has a sufficient sample size (i.e., high power) is common practice and an expected requirement for funding and/or publication, while in others it rarely occurs.  For example, \cite{charles2009reporting} found that 95\% of randomized controlled trials report sample size calculations.  In contrast, only a tiny fraction, estimated at about 3\%, of psychological articles report statistical power \citep{fritz2013comprehensive}, and in conservation biology, the number is only marginally higher at approximately 8\% \citep{fidler2006impact}.

One argument is that, once a significant finding is achieved, the size of a study is no longer relevant.  \cite{aycaguer2013explicacion} explain as follows: ``If a study finds important information by blind luck instead of good planning, I still want to know the results.''   Another viewpoint is that, while far from ideal, underpowered studies should be published since cumulatively, they can contribute to useful findings \citep{walker1995low}.  Others, disagree and contend that small sample sizes undermine the reliability of published science \citep{button2013empirical, dumas2017low, nord2017power}. In the context of clinical trials, \cite{inthout2016obtaining} review the many conflicting opinions about whether trials with suboptimal power are justified and conclude that, in circumstances when evidence for efficacy can be effectively combined across a series of trials (e.g. via meta-analysis), small sample sizes might be justified.

Despite the long-running and ongoing debates on significance thresholds and sample size requirements, there has been little to no modeling of how changes to a publication policy might affect what type of studies are pursued, the incentive structures driving research, and ultimately, the reliability of published science. One example is \cite{borm2009publication} who conclude, based on simulation studies, that the negative impact of publication bias does not warrant the exclusion of trials with low power.  Another recent example is \cite{higginson2016current}, who, based on results from an optimality model of the ``scientific ecosystem'', conclude that in order to ``improve the scientific value of research'' peer-reviewed publications should indeed require larger sample sizes, lower the $\alpha$ threshold, and give ``less weight to strikingly novel findings''.  Our work here aims to build on upon these modeling efforts to better inform the ongoing discussion on the reproducibility of published science.

\begin{figure}
\begin{center}
\includegraphics[width=6in]{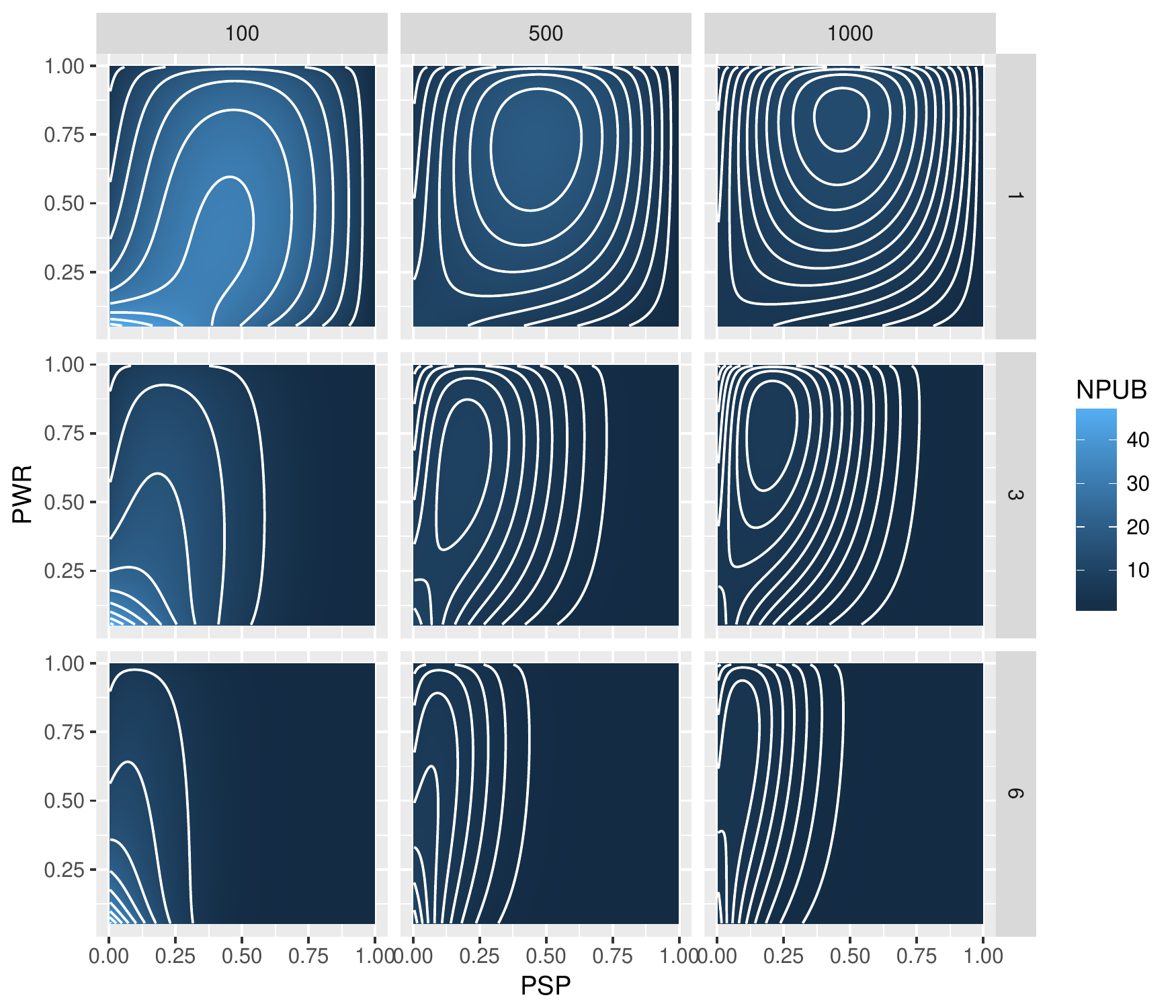}
\end{center}
\caption{Plots show the number of expected publications, $N_{PUB}$, for different values of $PSP$ and $PWR$.  Each panel represents one ecosystem defined by $\alpha=0.05$, $A = (1-psp)^{m},$ and $B=0$, with $k$ and $m$ as indicated by column and row labels respectively.  Depending on $k$ and $m$, the value of ($PSP$, $PWR$) that maximizes $N_{PUB}$ can change substantially.  With a higher $k$, higher-powered strategies will yield a greater $N_{PUB}$; with smaller $m$, optimal strategies are those with higher $PSP$.  Assuming that researchers behave based on optimizing thee use of their resources, it is interesting to observe how the``optimal strategy'' changes under different scenarios. \label{fig:heat}}
\end{figure}

 This paper is structured as follows.  In Section 2, we describe the model and methodology proposed to evaluate different publication policies.  We also list a number of metrics of interest.  In Section 3, we use the proposed methodology to evaluate potential effects of lowering the significance threshold; and in Section 4, the effects of requiring larger sample sizes.  Finally, in Section 5, we conclude with suggestions as to how publication policies can be defined to best improve the reliability of published research.

\section{Methods}
\label{sec:meth}

Recently, economic models have been rather useful for evaluating proposed research reforms \citep{gall2017credibility}.   However, modeling of how resources ought to be allocated among research projects is not new.  See for example, the work of \cite{greenwald1975consequences}, \cite{dasgupta1987simple}, and \cite{mclaughlin2011pursuit}.  Our framework for modeling the scientific ecosystem is closest in spirit to that of \cite{higginson2016current} who formulate a relationship between a researcher's strategy and his/her payoff, with the strategy involving a choice mix between exploratory and confirmatory studies, and a choice of pursuing fewer studies with larger samples or more studies with smaller samples. In Section 2.3, we comment in detail on the similarities and differences between our approach and that of \cite{higginson2016current}.

The publication process is complex and includes both objective and subjective considerations of efficacy and relevance.  The title of this article was chosen specifically to emphasize this point \citep{gornitzki2015freewheelin}.  A large, complicated human process like that of scientific publication cannot be entirely reduced to metrics and numbers: there are often financial, political and even cultural reasons for a paper being accepted or rejected for publication. With this in mind, the model presented here should not be seen as an attempt to precisely map out the peer-review process, but rather, as a useful tool for determining the consequences of implementing different publication policies.  

Within our optimality model, many assumptions and simplifications are made.  Most importantly, we assume that each researcher must make decisions consisting of only two choices: what statistical power (i.e., sample size) to adopt and what \textcolor{black}{``pre-study probability''} to pursue. Before elaborating further, let us briefly discuss these two concepts.

\subsection{Statistical power}

Increasing statistical power by conducting studies with larger sample sizes would undeniably result in more published research being true. However, these improvements may only prove modest, given current publication guidelines.  When we consider the perspectives of both researchers and journal editors, it is not surprising that statistical power has not improved \citep{smaldino2016natural} despite being highlighted as an issue over five decades ago \citep{cohen1962statistical}.

From a researcher's perspective, there is little incentive to conduct high-powered studies: basic logic suggests that the likelihood of publication is only minimally affected by power. To illustrate, consider a large number of hypotheses tested, out of which 10\% are truly non null.  Under the assumption that only (and all) positive results are published with $\alpha=0.05$ (which may in fact be realistic in certain fields, \cite{fanelli2011negative}), increasing average power from an ``unacceptably low'' 55\% to a ``respectable'' 85\% (at the cost of more than doubling sample size), results in only a minimal increase in the likelihood of publication:  from 10\% to 13\%.  Moreover, the proportion of true findings amongst those published is only increased modestly: from 55\% to 64\%.  Indeed, a main finding of \cite{higginson2016current} is that the rational strategy of a researcher is to ``carry out lots of underpowered small studies to maximize their number of publications, even though this means around half will be false positives.''  This result is in line with the views of many; see for example  \cite{bakker2012rules}, \cite{button2013power} and \cite{gervais2015powerful}.

From a journal's perspective, there is also little incentive to require larger sample sizes as a requirement for publication.  \cite{fraley2014n} review the publication history of six major journals in social-personality psychology and find that ``journals that have the highest impact [factor] also tend to publish studies that have smaller samples''.   This finding is in agreement with \cite{szucs2016empirical} who conclude that, in the fields of cognitive neuroscience and psychology, journal impact factors are negatively correlated with statistical power; see also \citet{brembs2013deep}.

\subsection{Pre-study probability}

We use the term ``pre-study probability'' ($psp$) as shorthand for the a-priori probability that a study's null hypothesis is false.  In this sense, highly exploratory research will typically have very low $psp$, whereas confirmatory studies will have a relatively high $psp$.  Studies with low $psp$ are not problematic per se.  To the contrary, there are undeniable benefits to pursuing ``long-shot'' novel ideas that are very unlikely to work out, see \cite{cohen2017should}.  While replication studies (i.e., studies with higher $psp$) may be useful to a certain extent, there is little benefit in confirming a result that is already widely accepted.  As \cite{button2013empirical} note: ``As R [the pre-study odds] increases [...] the incremental value of further research decreases.''   Most scientific journals no doubt take this into account in deciding what to publish, with more surprising results more likely to be published.  In turn, researchers deciding which hypotheses to pursue towards publication will emphasize those with lower $psp$.

Recognize that the lower the $psp$, the less likely a ``statistically significant'' finding is to be true.  As such, we are bound to a ``seemingly inescapable trade-off'' \citep{fiedler2017constitutes} between the novel and the reliable.  Journal editors face a difficult choice.  Either publish studies that are surprising and exciting yet most likely false, or publish reliable studies which do little to advance our knowledge.  Based on a belief that both ends of this spectrum are equally valuable,  \cite{higginson2016current} conclude that, in order to increase reliability, current incentive structures should be redesigned, ``giving less weight to strikingly novel findings.'' This is in agreement with the view of \cite{ten2016novel} who writes: ``If we are truly concerned about scientific reproducibility, then we need to reexamine the current emphasis on novelty and its role in the scientific process.''


\subsection{Model Framework}

The model and methodology we present seeks to add three features absent from the model of \cite{higginson2016current}.  While these authors consider how researchers balance available resources between exploratory and confirmatory studies, this simple dichotomy does not allow for a detailed assessment of the willingness of researchers to pursue high-risk studies.  Secondly, \cite{higginson2016current} define the ``total fitness of a researcher'' (i.e., the payoff for a given research strategy) with diminishing returns for confirmatory studies, but not for exploratory studies.  This choice, however well intended, has problematic repercussions for their optimality model.  (Under their framework, the optimal research strategy will depend on $T$, an arbitrary total budget parameter.)  Finally, by failing to incorporate the number of correct studies that go unpublished within their metric for the value of scientific research, many potential downsides of adopting measures to increase statistical stringency are ignored.  Other differences between our approach and previous ones will be made evident and include: considering  outcomes in terms of distributional differences, and specific modeling of how sample size requirements are implemented.

We describe our model framework in 5 simple steps.

(1) We assume, for simplicity, that all studies test a null hypothesis of equal population means against a two-sided alternative, with a standard two-sample Student $t$-test.  Each study has an equal number of observations per sample ($n_{1}=n_{2}$; $n=n_{1}+n_{2}$).  Furthermore, let us assume that a study can have one of only two results: (1) \emph{positive} ($p$-value $\le \alpha$), or (2) \emph{negative} ($p$-value $> \alpha$).  Given the true effect size, $\mu_{d}=\mu_{2}-\mu_{1}$ (the difference in population means), and $\sigma^{2}$, the common variance of each observation, we can easily calculate the probability of each result, using the standard formula for power. The probability of a positive result is equal to:

$Pr(positive; \mu_{d}, \sigma) = (1-F_{n-2, \frac{\mu_{d}}{\sigma^{*}}} ( t^{*}_{\alpha/2} ) ) + F_{n-2, \frac{\mu_{d}}{\sigma^{*}}} (-t^{*}_{\alpha/2} ))$, \\

\noindent where $t^{*}_{\alpha/2}$ is the upper 100$\cdot{\frac{\alpha}{2}}$-th percentile of the $t$-distribution with $n-2$ degrees of freedom,  $\sigma^{*}=\sigma\sqrt{(1/n_{1} +1/n_{2})}$, and $F_{df, ncp}(x)$ is the cdf of the non-central $t$ distribution with $df$ degrees of freedom and non-centrality parameter $ncp$.  For negative results we have $Pr(negative)=1- Pr(positive)$.  Then, for a given effect size $\delta$, we have the probability of a True Positive (TP), False Negative (FN), False Positive (FP), and True Negative (TN), equal to: $ Pr(TP) = Pr(Positive | \mu_{d} = \delta)$,   $ Pr(FN) = Pr(Negative | \mu_{d} = \delta) $ , $ Pr(FP) = Pr(Positive | \mu_{d} = 0)$ , and $ Pr(TN) = Pr(Negative | \mu_{d} = 0)$, respectively.

(2) Next we consider a large number of studies, $n_{S}$, each with a total sample size of $n$.  Of these $n_{S}$ studies, only a fraction, $psp$ (where $psp$ is the pre-study probability), have a true effect size of $\mu_{d} = \delta$. For the remaining $(1- psp)\cdot n_{S}$ studies, we have $\mu_{d} = 0$.  Throughout this paper, we keep $\delta=0.21$ and $\sigma^{2}=1$, as in \cite{higginson2016current}. (See their paper and the references within for a discussion of typical effect sizes across the psychology literature.)  Note that for a given sample size, $n$, these $n_{S}$ studies are each ``powered'' at level $pwr= Pr(TP)$.

\begin{table}
\caption{Equations for the expected number of studies (out of a total of $n_{S}$ studies) for each of the eight categories; with $A$ = prob. of publication for a positive result and  $B$ = prob. of publication for a negative result. \label{tab:equations}}
\begin{center}
\begin{tabular}{ l | l }
  \hline	
\textbf{Expected Number of...} & \textbf{Equation}\\\hline
 True Positives published & $TP_{PUB}=psp\cdot n_{S}\cdot A \cdot Pr(TP)$  \\
True Positives unpublished & $TP_{UN}=psp\cdot n_{S}\cdot (1-A) \cdot Pr(TP)$  \\
False Nulls published  & $FN_{PUB}=psp\cdot n_{S}\cdot B \cdot  Pr(FN) $ \\
False Nulls unpublished & $FN_{UN}=psp\cdot n_{S}\cdot (1-B) \cdot Pr(FN) $ \\
 False Positives published  & $FP_{PUB}=(1-psp)\cdot n_{S}\cdot A \cdot Pr(FP) $ \\
  False Positives unpublished & $FP_{UN}=(1-psp)\cdot n_{S}\cdot (1-A)\cdot Pr(FP)$ \\
  True Nulls published  & $TN_{PUB}=(1-psp)\cdot n_{S}\cdot B \cdot Pr(TN)$  \\
  True Nulls unpublished  & $TN_{UN}=(1-psp)\cdot n_{S}\cdot (1-B) \cdot Pr(TN) $  \\
   \hline
\end{tabular}
\end{center}
\end{table}

(3) We also label each study as either published (PUB) or unpublished (UN) for a total of 8 distinct categories (= 2 (positive, null) x 2 (true and false) x 2 (published and unpublished)).  One can determine the expected number of studies (out of a total of $n_{S}$ studies) in each category  by simple arithmetic.  Table 1 lists the equations for each of the eight categories with $A$ equal to the probability of publication for a positive result, and $B$ equal to the probability of publication for a negative result.  The parameters $A$ and $B$ may be fixed numbers (e.g. $A$=1, $B$=0.1, representing a scenario in which all positive results are published, and 10\% of null results are published), or defined as functions of study characteristics (e.g. $A =  (1-psp)^{m}$, a decreasing function of $psp$, representing a scenario in which positive results with lower $psp$ are more likely to be published on the basis of novelty).

(4) We determine the total number of studies, $n_{S}$, based on three parameters:  $T$, the total resources available (in units of observations sampled); $k$, the fixed cost per study (also expressed in equivalent units of observations sampled); and $n$, the total sample size per study.  Consequently, as in \cite{higginson2016current}, $n_{S} = {(k+n)^{-1}}{T}$.  Then, for any given level of power,  $pwr$, we have a necessary sample size per study, $n$, and a resulting total number of studies, $n_{S}$.  Throughout this paper, when necessary, we take $T=100,000$.  However, note that when comparing the outcomes of different publication policies, this choice is entirely irrelevant.

(5) Finally, let us define a ``research strategy'' to be a given pair of values for ($psp$, $pwr$) within ([0,1] x [$\alpha$,1]).  Then, for a given research strategy we can easily calculate the total expected number of publications:
\begin{equation}
N_{PUB}(psp, pwr) = TP_{PUB} + FN_{PUB} + FP_{PUB} + TN_{PUB}.
\end{equation}

With this setup in hand, suppose now a researcher pursues -consciously or unconsciously- strategies that maximize the expected number of publications \citep{charlton2006should}.  (This may not be an entirely unreasonable assumption, see \cite{van2014publication}.)  Figure \ref{fig:heat} shows how $N_{PUB}$ changes over a range of values of $psp$ and $pwr$ and under different fixed values for $k$ and $m$.  Depending on $k$ and $m$, the value of ($psp$, $pwr$) that maximizes $N_{PUB}$ can change substantially.  With larger $k$, higher-powered strategies will yield a greater $N_{PUB}$; with smaller $m$, optimal strategies are those with higher $psp$.  It is interesting to observe how the optimal strategy changes under different scenarios.  However, it may be more informative to consider a \emph{distribution} of preferred strategies.  This may also be a more realistic approach.  While rational researchers may be drawn toward optimal strategies, surely scientists are not willing and/or able to precisely identify these.  

Let us introduce some compact notation that will be useful for expressing distributional quantities of interest.  Particularly, the probabilities comprising the distribution of a study across the eight categories are expressed as $q_{abc}(psp, pwr)$, where $a \in \{0,1\}$ indicates the truth ($a=0$ for null, $a=1$ for alternative), $b \in \{0,1\}$ indicates the statistical finding ($b=0$ for negative, $b=1$ for positive),
and $c \in \{U,P\}$ indicates publication status.
As examples, we could write $FN_{UN} = q_{10U}$, or $TP_{PUB} = q_{11P}$.
We also use a plus notation to add over subscripts, so,
for instance, $q_{1+P} = TP_{PUB} + FN_{PUB}$.

As motivated above, we consider properties that result from a scientist or group of scientists stochastically allocating $T$ {\em resources} (not studies per se) according to a distribution across $(psp,pwr)$.   Presuming the incentive to publish, the density of this distribution is taken proportional to $N_{PUB}(psp,pwr)$, which we express as
\begin{eqnarray}
f_{RES}(psp, pwr) & \propto & N_{PUB}(psp,pwr) \nonumber \\
& \propto &   \{ k + n(pwr) \}^{-1} q_{++P}(psp, pwr). \label{f_res}
\end{eqnarray}
Consequently,
the distribution of $(psp, pwr)$  across {\em attempted}  studies has density
\begin{eqnarray}
f_{ATM}(psp, pwr) & \propto & \{ k + n(pwr) \}^{-1} f_{RES}(psp, pwr) \nonumber \\
& \propto & \{ k + n(pwr) \}^{-2} q_{++P}(psp, pwr). \label{f_atm}
\end{eqnarray}
In turn,
the distribution of $(psp,pwr)$ across {\em published studies} has density
\begin{eqnarray}
f_{PUB}(psp, pwr) & \propto & f_{ATM}(psp,pwr) q_{++P}(psp,pwr) \nonumber \\
& \propto &  \{ k + n(pwr) \}^{-2} \{ q_{++P}(psp, pwr) \}^{2}. \label{f_pub}
\end{eqnarray}
Note particularly that $f_{PUB}(psp,pwr) \propto \{f_{RES}(psp,pwr)\}^{2}$.
Hence the distribution of $(psp,pwr)$ across published studies is a concentrated version of the distribution describing how resources are deployed.

Armed with (\ref{f_res}), (\ref{f_atm}), and (\ref{f_pub}), we can investigate  the properties of a given scientific ecosystem, and how these properties vary across ecosystems.
Specifically, an ecosystem is specified by choices of $\alpha$, $k$, $m$, $A$ and $B$.
For any specification, properties of the three distributions are readily computed via 
two-dimensional numerical integration using a fine grid of $(psp,pwr)$ values.

\subsection{Ecosystem Metrics}

We will evaluate each ecosystem of interest on the basis of the following six metrics. 

\subsubsection{Reliability}

A highly relevant metric for the scientific ecosystem is the proportion of published findings that are correct.
In all the ecosystems we consider in this paper, we make the assumption that only positive results are published (i.e., $B=0$).  Therefore, we can express reliability (REL) simply as:
\begin{eqnarray*}
REL &=& \frac
{ E_{ATM}
\left\{  q_{11P}(PSP, PWR) \right\}
}
{ E_{ATM}
\left\{  q_{+1P}(PSP, PWR) \right\}
}.
\end{eqnarray*}

More generally, in ecosystems where negative results might be published (i.e., $B(psp,pwr)\ne0$), 
the reliability would equal the proportion of published papers that reach a correct conclusion, i.e.,
\begin{eqnarray*}
REL &=& \frac
{ E_{ATM}
\left\{  q_{00P}(PSP, PWR) + q_{11P}(PSP, PWR) \right\}
}
{ E_{ATM}
\left\{  q_{++P}(PSP, PWR) \right\}
}.
\end{eqnarray*}

\subsubsection{Number of Studies Attempted/Published}

If $T$ units of resources are deployed according to $f_{RES}()$,
then we expect that $N_{ATM}$ studies will be attempted, where
\begin{eqnarray*}
T^{-1}N_{ATM} &=& E_{RES} \left[ \{k+n(PWR)\}^{-1} \right].
\end{eqnarray*}
Similarly, we expect $N_{PUB}$ studies will be published, with
\begin{eqnarray*}
T^{-1}N_{PUB} &=& E_{RES} \left[ \{k+n(PWR)\}^{-1} q_{++P}(PSP,PWR)   \right].
\end{eqnarray*}
The ratio $N_{PUB}/N_{ATM}$, which does not depend on $T$, is of evident interest,
as the {\em publication rate} (PR) for attempted studies.

\subsubsection{Rate of Silenced True Positive Research}

Another quantity attached to an ecosystem is the fraction of true positives that end up unpublished.   This {\em silenced true positive rate} (STPR) is given by
\begin{eqnarray*}
STPR &=&
\frac{ E_{ATM} \{ q_{11U}(PSP, PWR)\}}
{E_{ATM}\{ q_{11+}(PSP, PWR)\}}.
\end{eqnarray*}

\subsubsection{Balance between Exploration and Confirmation}

There has been much discussion about the desired balance between researchers looking for {\em a priori} unlikely relationships versus confirming suspected relationships put forth by other researchers; see for example, \cite{sakaluk2016exploring} and \cite{kimmelman2014distinguishing}.  The marginal distribution of $PSP$ arising from $f_{ATM}(psp,pwr)$ describes the balance between exploration versus confirmation for attempted studies, while the $PSP$ marginal from $f_{PUB}(psp,pwr)$ does the same for published studies.   More specifically, we report the interquartile ranges of these marginal distributions for a given ecosystem.

\subsubsection{Breakthrough Discoveries}

The ability of the scientific ecosystem to produce breakthrough findings is an important attribute.  We quantify this in terms of spending $T$ resource units yielding an expectation of $DSCV$ breakthrough results.  Here a breakthrough result is defined as a true positive and published study that results from a $psp$ value below a threshold, i.e., a very surprising positive finding that gets published and also happens to be true.  If we set the breakthrough threshold as $psp < 0.05$, then

\begin{eqnarray*}
T^{-1}DSCV  &=&
E_{RES} \left\{   I_{(0,0.05)}(PSP)  \frac{q_{11P}(PSP,PWR)} {k+n(PWR)} \right\}.
\end{eqnarray*}

\subsubsection{Power of Attempted/Published Studies}

We already mentioned the relevance of the $PSP$ marginals of (\ref{f_atm}) and (\ref{f_pub}). In a similar vein, the marginal distributions of $PWR$ under each of these distributions are readily interpreted metrics of the ecosystem.

\section{The effects of adopting lower significance thresholds}
\label{sec:loweralpha}

In this section, we investigate the impact of adopting lower significance thresholds.  Here we will assume that the sample size of a study does not affect the likelihood of publication and that studies with lower $psp$ are more likely to be published.  As such, we define: $A = (1-psp)^{m}$.  We will also assume that only positive studies are published, hence, $B = 0$.  We compute the metrics of interest for 36 different ecosystems.  Each ecosystem is uniquely defined with one of three possible values for $m$ (=1, 3, 6), one of three possible values for $k$ (=100, 500, 1000), and most importantly, one of four possible values for the $\alpha$ significance threshold (= 0.001, 0.005, 0.050, 0.10).   

\subsection{Results}
First, let us go over scenarios in which $\alpha$ is held fixed at 0.05 and $k$ and $m$ are varied, see Table \ref{tab:baseprop}.  We observe that, as $k$ increases, the reliability of published research decreases, as does the rate of breakthrough discoveries.  As $m$ increases, reliability decreases while the rate of breakthrough discoveries increases.  We should also note that the publication rate decreases with $m$ and increases with $k$.  While the PR numbers we obtain may appear rather low, consider that  \citet{siler2015measuring}, in a systematic review of manuscripts submitted to three leading medical journals, observed a publication rate of 6.2\%.  In a review of top psychology journals, \cite{lee2011social} found that rejection rates ranged between 68\% to 86\%.  

In Table \ref{compare0005to005}, we note how the various metrics change with an $\alpha=0.005$ relative to $\alpha=0.05$.  {Complete results are presented in Figures and Tables in the Appendix.}  Based on our results, we can make the following conclusions on the impact of adopting a lower, more stringent, significance threshold.

  \begin{table}[p]
\centering
\begin{tabular}{rrccc}
  \hline
 $k$ & $m$ & Publication Rate, & Reliability,  &   Breakthrough Discoveries, \\ 
  &  & ($PR = N_{PUB}:N_{ATM}$) & ($REL$) &     ($DSCV$)\\ 
  \hline
100 & 1 & 0.08 = \footnotesize{24.3 : 296.7} & \normalsize{0.76} & 0.10 \\ 
  100 & 3 & 0.04 = \footnotesize{13.8 : 317.6} & \normalsize{0.53} & 0.23 \\ 
  100 & 6 & 0.03 = \footnotesize{11.0 : 337.7} & \normalsize{0.35} & 0.41 \\ 
  500 & 1 & 0.11 = \footnotesize{12.5 : 112.1} & \normalsize{0.84} & 0.04 \\ 
  500 & 3 & 0.05 = \footnotesize{6.2 : 114.9} & \normalsize{0.66} & 0.11 \\ 
  500 & 6 & 0.04 = \footnotesize{4.4 : 117.8} & \normalsize{0.48} & 0.20 \\ 
  1000 & 1 & 0.12 = \footnotesize{8.4 : 68.7} & \normalsize{0.85} & 0.03 \\ 
  1000 & 3 & 0.06 = \footnotesize{4.1 : 69.7} & \normalsize{0.69} & 0.07 \\ 
  1000 & 6 & 0.04 = \footnotesize{2.8 : 70.7} & \normalsize{0.51} & 0.13 \\ 
   \hline
\end{tabular}
\caption{{For ecosystems defined by fixed $\alpha=0.05$, $A = (1-psp)^{m},$ and $B=0$, and varying values of $k$ and $m$, the table lists estimates for the publication rate, the reliability and the number of breakthrough discoveries per unit of resources.}}
\label{tab:baseprop}
\end{table}

\begin{table}[p]
\centering
\begin{tabular}{rrrrr}
  \hline
   $k$ & $m$ & Change in PR, & Change in REL,  &   Change in DSCV, \\ 
 &  & \footnotesize{($PR_{0.005} : PR_{0.05}$)}    & \footnotesize{($REL_{0.005}:REL_{0.05}$)}  & \footnotesize{($DSCV_{0.005}:DSCV_{0.05}$)} \\ 
  \hline
100 & 1 & \footnotesize{0.086 : 0.082} = \normalsize{1.05} & \footnotesize{0.982 : 0.761} = \normalsize{1.29} & \footnotesize{0.019 : 0.097} = \normalsize{ 0.20} \\ 
  100 & 3 & \footnotesize{0.035 : 0.043} = \normalsize{0.80} & \footnotesize{0.957 : 0.534} = \normalsize{1.79} & \footnotesize{0.053 : 0.229} = \normalsize{ 0.23} \\ 
  100 & 6 & \footnotesize{0.018 : 0.032} = \normalsize{0.56} & \footnotesize{0.919 : 0.349} = \normalsize{2.63} & \footnotesize{0.115 : 0.411} = \normalsize{ 0.28 }\\ 
  500 & 1 & \footnotesize{0.103 : 0.111} = \normalsize{0.93} & \footnotesize{0.985 : 0.837} = \normalsize{1.18} & \footnotesize{0.013 : 0.045} = \normalsize{ 0.29 }\\ 
  500 & 3 & \footnotesize{0.042 : 0.054} = \normalsize{0.77} & \footnotesize{0.965 : 0.656} = \normalsize{1.47} & \footnotesize{0.037 : 0.109} = \normalsize{ 0.34 }\\ 
  500 & 6 & \footnotesize{0.022 : 0.037} = \normalsize{0.59} & \footnotesize{0.934 : 0.475} = \normalsize{1.96} & \footnotesize{0.079 : 0.201} = \normalsize{ 0.39 }\\ 
  1000 & 1 & \footnotesize{0.112 : 0.122} = \normalsize{0.92} & \footnotesize{0.986 : 0.853} = \normalsize{1.16} & \footnotesize{0.010 : 0.029} = \normalsize{ 0.34 }\\ 
  1000 & 3 & \footnotesize{0.045 : 0.058} = \normalsize{0.77} & \footnotesize{0.968 : 0.687} = \normalsize{1.41} & \footnotesize{0.027 : 0.071} = \normalsize{ 0.38 }\\ 
  1000 & 6 & \footnotesize{0.024 : 0.039} = \normalsize{0.61} & \footnotesize{0.940 : 0.512} = \normalsize{1.84} & \footnotesize{0.059 : 0.132} = \normalsize{ 0.44 }\\ 
   \hline
\end{tabular}
\caption{{A comparison between ecosystems with $\alpha=0.05$ and ecosystems with $\alpha=0.005$.  For all we have $A = (1-psp)^{m}$, $B=0$, and varying values of $k$ and $m$.  The table lists estimates for the ratios of the publication rate (PR), the reliability (REL) and the number of breakthrough discoveries (DSCV) per unit of resources.}}
\label{compare0005to005}
\end{table}

 \begin{enumerate}
 \item{Reliability is substantially increased with a lower threshold. Based on our results, comparing $\alpha=0.005$ to $\alpha=0.05$, the increase in the probability that a published result is true ranges from a 16\% increase to a 163\% increase, see Table \ref{compare0005to005}.  The impact on REL is greatest when $k$ is small and $m$ is large.  This is due to the fact that with a lower significance threshold policy, attempted studies are typically of higher-power (particularly so when $k$ is small) and of higher pre-study probability, see Figures \ref{fig:meanpsp} and \ref{fig:medpwr} and Table \ref{tab:IQ005to05}.}
 \item{A disadvantage of the lower threshold is that the number of breakthrough discoveries is substantially lower.  Comparing results of $\alpha=0.005$ and $\alpha=0.05$, this ranges from a reduction of 56\% to a reduction of 80\% in the number of published positive studies with $psp \le 0.05$, see Table \ref{compare0005to005} and Figure \ref{fig:DSCV}.}
 \item{When sample size is less costly relative to the total cost of a study (i.e., when $k$ is larger), the benefit of lowering the significance threshold (increased $REL$) is somewhat smaller.  However, the downside (decreased in $DSCV$) is \emph{substantially} smaller, see left-panels of Figures \ref{fig:REL} and \ref{fig:DSCV}.  This suggests that a policy of lowering the significance threshold would perhaps be best suited in a field of research in which increasing one's sample size is less burdensome.  This nuance recalls the suggestion of \cite{ioannidis2013optimal}: ``Instead of trying to fit all studies to traditionally acceptable type I and type II errors, it may be preferable for investigators to select type I and type II error pairs that are optimal for the truly important outcomes and for clinically or biologically meaningful effect sizes.''}
 \item{ When novelty is more of a requirement for publication (i.e., when $m$ is larger), {the benefit of lowering the significance threshold is larger and the downside smaller.}  This result is due to the fact that a smaller $\alpha$ will incentivize researchers to allocate resources in the direction towards either higher-powered or higher $psp$ studies (i.e., away from the SW corner of the plots in Figure 1).  If a higher $psp$ more negatively impacts the chance that a study is published, then moving towards higher power (North) will be more favourable than towards higher $psp$ (East).  This suggests that for a lower significance threshold policy to be most effective, editors should also adopt, in conjunction, stricter requirements for research novelty.  To illustrate, consider three ecosystems of potential interest with their estimated $REL$, $DSCV$ and $PR$ metrics:\\
 (1) The \emph{baseline} defined by $\alpha=0.05$, $m=3$, and $k=500$ with:\\
 $REL=0.656$, $DSCV=0.109$, and $PR = 0.054$;\\
 (2) the \emph{alternative} defined by $\alpha=0.005$, $m=3$, and $k=500$ with:\\
  $REL=0.965$, $DSCV=0.037$, and $PR$ =0.042; and\\
 (3) the \emph{suggested}  defined by $\alpha=0.005$, $m=6$, and $k=500$ with:\\
  $REL=0.934$, $DSCV=0.079$, and $PR$ =0.022.\\
{Note that the while the \emph{suggested} has high REL and relatively high DSCV, the PR is substantially reduced.}
}
 \item{As expected, lowering the significance threshold has the effect of increasing the amount silenced true positive research ($STPR$), see Figure \ref{fig:silenced} (left-hand panel). The effect of lowering the significance threshold on STPR is approximately the same regardless of whether novelty is highly valued ($m$), and regardless of whether increasing sample size is expensive ($k$).}
 \item{As mentioned earlier, the balance between exploratory and confirmatory research is an important aspect of a scientific ecosystem.  The results show that the width of the IQ range for $psp$ does not change substantially with $\alpha$,  see Figure \ref{fig:IQpsp}.  As such, we could conclude that even with a much lower significance threshold, there will still be a wide range of studies attempted in terms their $psp$.  However, $psp$ values do tend to be substantially higher with smaller $\alpha$, see Table \ref{tab:IQ005to05}. As such we should expect that, with smaller $\alpha$, research will move towards more confirmatory, and less exploratory studies.}
  \end{enumerate}

\section{The effects of strict a-priori power calculation requirements}
\label{sec:power}

In this section, we investigate the effects of requiring larger sample sizes.  In practical terms, this means adopting publication policies that require studies to show a-priori power calculations indicating that sample sizes are ``sufficiently large'' to achieve the desired level of statistical power.
\textcolor{black}{
Whereas before, the chance of publishing a positive study in our framework depended only on novelty via
\begin{eqnarray*}
A &=& (1-psp)^{m},
\end{eqnarray*}
a convenient choice to represent a journal policy of requiring an a-priori sample size justification would be:
\begin{eqnarray}
A &=& \frac{(1-psp)^{m}}
{1+\exp\left\{ -(\log 19) \frac{pwr-c_{50}}{c_{95}-c_{50}} \right\}}.
\label{Apwr}
\end{eqnarray}
So $A$ is reduced more when power is lower,
with the extent of the reduction parameterized by $(c_{50},c_{95})$.
Specifically,
$c_{95}$ is the value for $pwr$ at which the multiplicative reduction is near negligible (factor of $0.95$), while $c_{50}$ is the value for $pwr$ at which the multiplicative reduction is a factor of $0.5$.
We conduct our experimentation using $c_{50},c_{95}=(0.5, 0.8)$, with the following rationale.
If a journal does require an a priori sample size justification,
a claim of 80\% power is the typical requirement.
Hence a study which really attains 80\% power is not likely to suffer in its quest for publication,
motivating $c_{95}=0.8$.
However, it is well commented upon \citep{bland2009tyranny, vasishth2017illusion} that often a-priori sample size claims are exaggerated through various mechanisms, meaning that a study with less than 80\% power might be advertised as having 80\% power.
This is the basis for setting $c_{50}=0.5$, i.e., truly possessing only 50\% power does substantially reduce, but not eliminate, the chance of publication.
} We calculated the metrics of interest for the same 36 different ecosystems as in Section 3, with the only difference being that the parameter $A$ is defined according to equation \ref{Apwr}.  To contrast these ecosystems with those discussed in the previous section, we refer to these ecosystems as ``with SSR'' (sample size requirements).

\subsection{Results}

\begin{table}[ht]
\centering
\begin{tabular}{rrrrr}
  \hline
 $k$ & $m$ & Change in PR, & Change in REL,  &   Change in DSCV, \\ 
 &  & \footnotesize{($PR_{SSR} : PR_{\hcancel[0.5pt]{SSR}}$)}    & \footnotesize{($REL_{SSR}:REL_{\hcancel[0.5pt]{SSR}}$)}  & \footnotesize{($DSCV_{SSR}:DSCV_{\hcancel[0.5pt]{SSR}}$)} \\ 
  \hline
100 & 1 & \footnotesize{0.119  : 0.082} = \normalsize{ 1.45} &  \footnotesize{0.899 : 0.761 } = \normalsize{ 1.18 }& \footnotesize{0.050 : 0.097} = \normalsize{ 0.52} \\ 
  100 & 3 & \footnotesize{0.054 : 0.043} = \normalsize{ 1.24} &  \footnotesize{0.780 : 0.534 } = \normalsize{ 1.46 }& \footnotesize{0.129 : 0.229} = \normalsize{ 0.56} \\ 
  100 & 6 & \footnotesize{0.034 : 0.032} = \normalsize{ 1.04} &  \footnotesize{0.639 : 0.349 } = \normalsize{ 1.83 }& \footnotesize{0.252 : 0.411} = \normalsize{ 0.61} \\ 
  500 & 1 & \footnotesize{0.132 : 0.111} = \normalsize{ 1.19} &  \footnotesize{0.902 : 0.837 } = \normalsize{ 1.08 }& \footnotesize{0.034 : 0.045} = \normalsize{ 0.76} \\ 
  500 & 3 & \footnotesize{0.060 : 0.054} = \normalsize{ 1.11} &  \footnotesize{0.785 : 0.656 } = \normalsize{ 1.20 }& \footnotesize{0.087 : 0.109 }= \normalsize{ 0.80} \\ 
  500 & 6 & \footnotesize{0.038 : 0.037} = \normalsize{ 1.01} &  \footnotesize{0.646 : 0.475 } = \normalsize{ 1.36} & \footnotesize{0.171 : 0.201 }= \normalsize{ 0.85} \\ 
  1000 & 1 & \footnotesize{0.139 : 0.122} = \normalsize{ 1.14} & \footnotesize{ 0.903 : 0.853 } = \normalsize{ 1.06} & \footnotesize{0.024 : 0.029 }= \normalsize{ 0.84} \\ 
  1000 & 3 & \footnotesize{0.063 : 0.058} = \normalsize{ 1.08} &  \footnotesize{0.788 : 0.687 } = \normalsize{ 1.15} & \footnotesize{0.063 : 0.071} = \normalsize{ 0.88 }\\ 
  1000 & 6 & \footnotesize{0.039 : 0.039} = \normalsize{ 1.00} &  \footnotesize{0.651 : 0.512 } = \normalsize{ 1.27} & \footnotesize{0.123 : 0.132} = \normalsize{ 0.93} \\ 
   \hline
\end{tabular}
\caption{{A comparison between ecosystems with a sample size requirement (SSR) and ecosystems without (\hcancel[1pt]{SSR}).  $A$ is defined by equation \ref{Apwr} (with $c_{50}=0.5$ and $c_{95}=0.8$), $B=0$, and varying values of $k$ and $m$.  The table lists estimates for the ratios of the publication rate (PR), the reliability (REL) and the number of breakthrough discoveries (DSCV) per unit of resources.}}
\label{tab:compareSSRtonoSSR}
\end{table}

Based on our results, we can make the following main conclusions on the measurable consequences of adopting a journal policy requiring an a-priori sample size justification.  

\begin{figure}[h]
\fbox{\includegraphics[width=12cm]{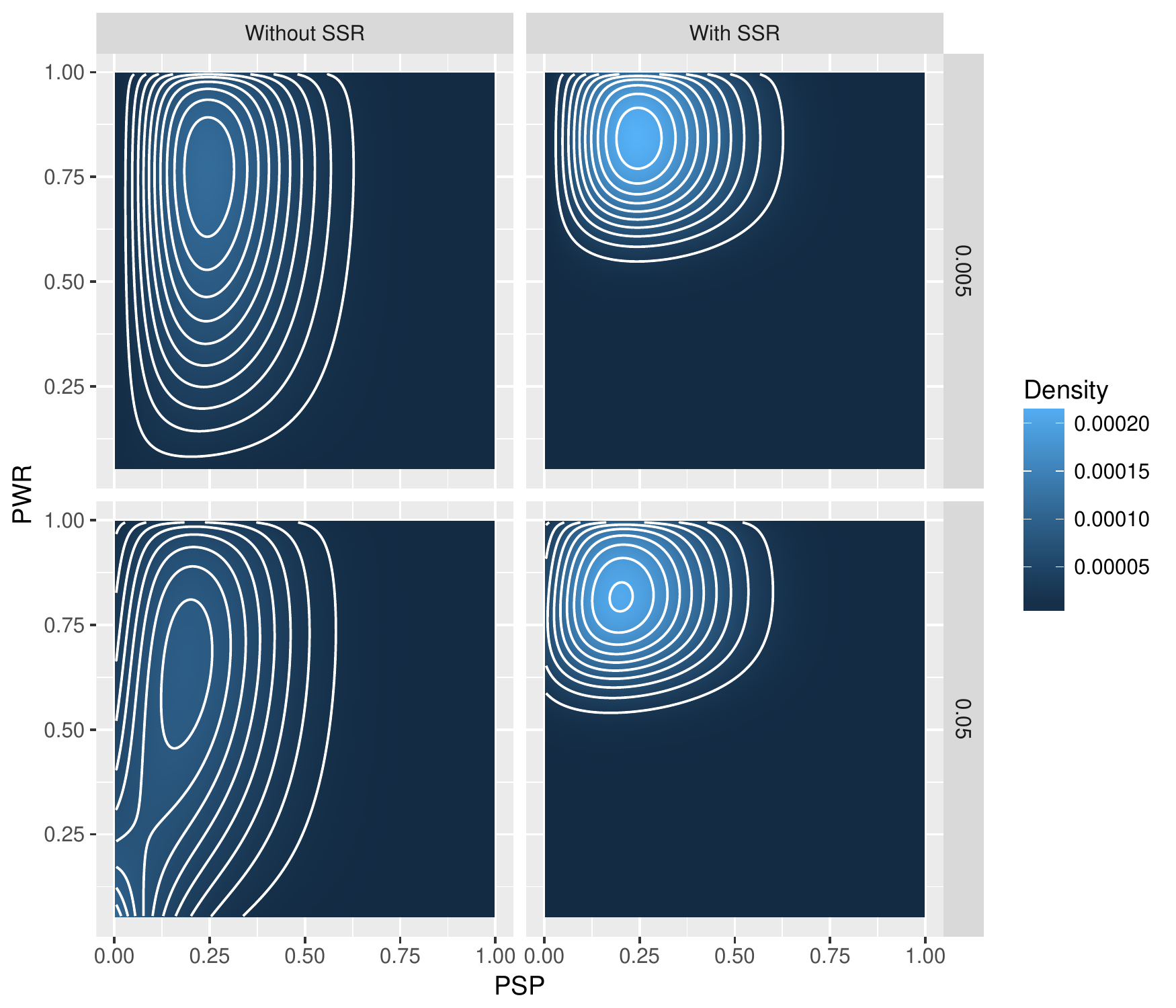}}
\caption{{Heat-maps show the density of published papers for the four policies of interest, with $k=500$ and $m=3$. 
} }
\label{fig:heat2}
\end{figure}

\begin{enumerate}
\item{With SSR, we observed much higher powered studies, see Figure \ref{fig:medpwr}.  The median $pwr$ amongst attempted studies with SSR (fixed $\alpha=0.05$) ranged between 0.705 and 0.770; and amongst published studies with SSR, the median $pwr$ ranged between  0.775 and 0.825.}
\item{The impact of requiring ``sufficient'' sample sizes, with regards to reliability, is similar to the impact of lowering the significance threshold: reliability is improved and particularly so when novelty is highly prized ($m$ is large).  With SSR, it is interesting to see that reliability is the same regardless of $k$, see Figure \ref{fig:REL}.  This can be explained by the fact that, in deciding on a sample size, cost will no longer be as much of a consideration with SSR.  Whereas the gains made in REL due to lowering $\alpha$ were a result of both higher $psp$ and $pwr$ studies, the increased REL in studies with SSR, is due primarily to increased $pwr$; see Figures \ref{fig:heat2}, \ref{fig:meanpsp} and  \ref{fig:medpwr}. }
\item{With regards to the amount of breakthrough discoveries ($DSCV$), the impact of having a minimum power requirement as a requisite for publication is greatest when both $k$ and $m$ are small.  This reduction in $DSCV$ can be substantial.  See Table \ref{tab:compareSSRtonoSSR} to compare ecosystems with SSR and without SSR (with fixed $\alpha=0.05$), and varying $k$ and $m$.  The decline in $DSCV$ ranges from 7\% to 48\%.}
\item{In conjunction with requiring larger sample sizes, it may be wise to place greater emphasis on research novelty.  As in Section 3, such a combined approach could see an increase in reliability with only a limited decrease in discovery.  This trade-off is most beneficial when $k$ is small.  Consider three ecosystems of potential interest (all with $\alpha=0.05$) with their estimated $REL$, $DSCV$ and $PR$ metrics:\\
 (1) The \emph{baseline} defined by $m=3$, $k=100$, and without SSR; with $REL=0.534$, $DSCV=0.229$,  $PR=0.043$; \\
 (2) the \emph{alternative} defined by $m=3$, $k=100$ and with SSR;  with $REL=0.780$, $DSCV=0.129$,  $PR=0.054$; and\\
 (3) the \emph{suggested}  defined by $m=6$, $k=100$ and with SSR; with $REL=0.639$, $DSCV=0.252$,  $PR=0.034$.\\
 {Note that the while the \emph{suggested} has both higher REL and higher $DSCV$ than the \emph{baseline}, the PR is reduced.}
}
\item{There is a substantial increase in STPR amongst ecosystems with SSR; see Figure \ref{fig:silenced}. With a minimum power requirement as a requisite for publication, there will be many more true positive findings that are not published.}
\item{The estimated inter-quantile range of $psp$ values (both attempted and published) is relatively unchanged by the sample size requirement, see Figure \ref{fig:IQpsp}.}
  \end{enumerate}

\subsection{In tandem: The effects of adopting both a lower significance threshold and a power requirement}

We are also curious as to whether lowering the significance threshold \emph{in addition} to requiring larger sample sizes would carry any additional benefits relative to each policy innovation on its own.  On this, we have the following results:
%

\begin{enumerate}
\item{For ecosystems with SSR, the distribution of published studies does not change dramatically when $\alpha$ is lowered.  Figure \ref{fig:heat2} shows the density over published papers for four ecosystems (all with $m=3$ and $k=500$):  (1) $\alpha=0.05$, without SSR; (2) $\alpha=0.05$, with SSR; (3) $\alpha=0.005$, without SSR; and (4) $\alpha=0.005$, with SSR. The difference between the densities of (2) and (4) is primarily a matter of a shift in $psp$.}
\item{As expected, lowering the significance threshold further increases reliability and the number of breakthrough discoveries is further decreased; see Figures \ref{fig:REL} and \ref{fig:DSCV}. Consider for example, ecosystems with fixed $m=1$ and $k=100$. Changing from a policy without SSR and $\alpha=0.05$, to a policy with SSR and $\alpha=0.005$, leads to a 30\% increase in reliability, a 85\% decrease in breakthrough discoveries. The publication rate (PR) is increased by 45\%, see Table \ref{tab:compareSSR005tonoSSR05}.  Note that, while for such a policy change, the PR increases, the number of publications actually decreases quite substantially, see Figure \ref{fig:Npub}.}
\item{For ecosystems with SSR, the number of publications, $N_{PUB}$, decreases with decreasing $\alpha$.  However, the rate of decline is much lower than the equivalent decline for ecosystems without SSR, see Figure \ref{fig:Npub}.  For ecosystems with SSR, the STPR does not decrease any further as $\alpha$ is lowered; see Figure \ref{fig:silenced}.}
  \end{enumerate}

\begin{table}[ht]
\centering
\begin{tabular}{rrrrr}
  \hline
 $k$ & $m$ & Change in PR, & Change in REL,  &   Change in DSCV, \\ 
 &  & \footnotesize{($PR_{\frac{SSR}{0.005}} : PR_{\frac{\hcancel[0.5pt]{SSR}}{0.05}}$)}    & \footnotesize{($REL_{\frac{SSR}{0.005}} : REL_{\frac{\hcancel[0.5pt]{SSR}}{0.05}}$)} & \footnotesize{($DSCV_{\frac{SSR}{0.005}}:DSCV_{\frac{\hcancel[0.5pt]{SSR}}{0.05}}$)} \\ \\
  \hline
100 & 1 & \footnotesize{0.119 : 0.082 } = \normalsize{ 1.45} & \footnotesize{0.990 : 0.761 } = \normalsize{ 1.30} & \footnotesize{0.015 : 0.097 } = \normalsize{ 0.15} \\ 
  100 & 3 & \footnotesize{0.048 : 0.043 } = \normalsize{ 1.10} & \footnotesize{0.977 : 0.534 } = \normalsize{ 1.83} & \footnotesize{0.042 : 0.229 } = \normalsize{ 0.18} \\ 
  100 & 6 & \footnotesize{0.025 : 0.032 } = \normalsize{ 0.78} & \footnotesize{0.958 : 0.349 } = \normalsize{ 2.74} & \footnotesize{0.091 : 0.411 } = \normalsize{ 0.22} \\ 
  500 & 1 & \footnotesize{0.126 : 0.111 } = \normalsize{ 1.13} & \footnotesize{0.990 : 0.837 } = \normalsize{ 1.18} & \footnotesize{0.011 : 0.045 } = \normalsize{ 0.26} \\ 
  500 & 3 & \footnotesize{0.051 : 0.054 } = \normalsize{ 0.93} & \footnotesize{0.978 : 0.656 } = \normalsize{ 1.49} & \footnotesize{0.032 : 0.109 } = \normalsize{ 0.30} \\ 
  500 & 6 & \footnotesize{0.027 : 0.037 } = \normalsize{ 0.71} & \footnotesize{0.959 : 0.475 } = \normalsize{ 2.02} & \footnotesize{0.071 : 0.201 } = \normalsize{ 0.35} \\ 
  1000 & 1 & \footnotesize{0.131 : 0.122 } = \normalsize{ 1.07} & \footnotesize{0.991 : 0.853 } = \normalsize{ 1.16} & \footnotesize{0.009 : 0.029 } = \normalsize{ 0.31} \\ 
  1000 & 3 & \footnotesize{0.053 : 0.058 } = \normalsize{ 0.90} & \footnotesize{0.978 : 0.687 } = \normalsize{ 1.42} & \footnotesize{0.025 : 0.071 } = \normalsize{ 0.36} \\ 
  1000 & 6 & \footnotesize{0.028 : 0.039 } = \normalsize{ 0.70} & \footnotesize{0.959 : 0.512 } = \normalsize{ 1.87} & \footnotesize{0.056 : 0.132 } = \normalsize{ 0.42} \\ 
   \hline
\end{tabular}
\caption{{A comparison between ecosystems with both sample size a requirement (SSR) and $\alpha=0.005$ and ecosystems without a sample size requirement (\hcancel[1pt]{SSR}) and $\alpha=0.05$.  For those with SSR, $A$ is defined by equation \ref{Apwr} (with $c_{50}=0.5$ and $c_{95}=0.8$).  For those without SSR, $A = (1-psp)^{m}$. For all we have $B=0$, and varying values of $k$ and $m$ as indicated.  The table lists estimates for the ratios of the publication rate (PR), the reliability (REL) and the number of breakthrough discoveries (DSCV) per unit of resources.}}
\label{tab:compareSSR005tonoSSR05}
\end{table}

\section{Conclusion}
\label{sec:conc}

There remains substantial disagreement on the merits of requiring ``greater statistical stringency'' to address the reproducibility crisis. Yet all should agree that innovative publication policies can be part of the solution.  Going forward, it is important to recognize that current norms for Type 1 and Type 2 error levels have been driven largely by tradition and inertia rather than careful coherent planning and result-driven decisions \citep{hubbard2003confusion}.  Hence, improvements should be possible.

In response to \cite{amrhein2017earth} who suggest that a more stringent $\alpha$ threshold will lead to published science being \emph{less} reliable, our results suggest otherwise.  However, just as \cite{amrhein2017earth} contend, our results indicate that the publication rate will end up being substantially lower with a smaller $\alpha$.  While going from $p<0.05$ to $p<0.005$ may be beneficial to published science in terms of reliability, we caution that there may be a large cost in terms of fewer breakthrough discoveries.  Importantly, and somewhat unexpectedly, our results suggest that this can be mitigated (to some degree) by adopting a greater emphasis on research novelty.  This approach however, might be difficult to achieve in practice, unless one is willing to accept a much lower publication rate.  In summary, publishing \emph{less} may be the necessary price to pay for obtaining \emph{more} reliable science.  

Recently, some have suggested that researchers choose (and justify) an ``optimal'' value for $\alpha$, for each unique study; see \cite{mudge2012setting}, \cite{ioannidis2013optimal} and \cite{lakens2018justify}.  Each study within a journal would thereby have a different set of criteria.  This is a most interesting idea and there are persuasive arguments in favor such an approach.  Still, it is difficult to anticipate how such a policy would play out in practice and how the research incentive structure would change in response.

We are also cautious about greater sample size requirements.  While improving reliability, requiring studies to show ``sufficient power'' will severely limit novel discoveries in fields where acquiring data is expensive.  In addition, a greater proportion of valid findings will be silenced (i.e., rejected for publication).  Is it beneficial for editors and reviewers to consider whether a study has ``sufficient power''?  How much should these criteria influence publication decisions?  Answers to these questions are not at all obvious.  Again, using the methodology introduced, we suggest that adopting a greater emphasis on research novelty may mitigate, to a certain extent, some of the downside of adopting greater sample size requirements at the cost of lowering the overall number of published studies.  {Given that, as a result of publication bias, it can often be better to discard 90\% of published results  for meta-analytical purposes \citep{stanley2010could}, this may be an approach worth considering.}

Our main recommendation is that, before adopting any (radical) policy changes, we should take a moment to carefully consider, and model how, these proposed changes might impact outcomes.  The methodology we present here can be easily extended to do just this.  Two scenarios of interest come immediately to mind.

First, it would be interesting to explore the impact of publication bias \citep{sterling1995publication}.  This could be done by allowing $B$ to take different non-zero values.  Based on simulation studies, \cite{de2013selective} suggest that publication bias can in fact be beneficial for the reliability of science.  However, under slightly different assumptions,  \cite{van2014publishing} arrive at very different conclusion.  Clearly, a better understanding of how publication bias changes a scientist's incentives is needed.

Second, it would be worthwhile to investigate the potential impact of requiring study pre-registration.  \cite{coffman2015pre} use the accounting of \cite{ioannidis2005most} to evaluate the effect of pre-registration on reliability and conclude that pre-registration will have only a modest impact.  However, the impact on the publication rate and on the number of breakthrough discoveries is still not well understood.  This is particularly relevant given the current trend to adopt ``result-blind peer-review'' \citep{greve2013result} policies including most recently, the policy of \emph{Registered Reports} \citep{chambers2015registered}.

Our methodology assumes above all that researchers' decisions are driven exclusively by the desire to publish.  But the situation is more complex.  Publication is not necessarily the end goal for a scientific study and requirements with regards to significance and power are not only encountered at the publication stage.  In the planning stages, before a study even begins, ethics committees and granting agencies will often have certain minimal requirements; see \cite{ploutz2014justifying} and \cite{halpern2002continuing}.  And after a study is published, regulatory bodies and policy makers will also often subject the results to a different set of norms.  

Finally, it is important to acknowledge that no publication policy will be perfect, and we must always be willing to accept that a certain proportion of research is potentially false \citep{djulbegovic2007should}.  Each policy will have its advantages and disadvantages.  Our modeling exercise makes this all the more evident and forces us to carefully consider different potential trade-offs.

\bibliography{truthinscience}

\pagebreak 

\begin{center}
{\large\bf SUPPLEMENTARY MATERIAL/APPENDIX}
\end{center}

\begin{description}

\item[Title:] Additional Tables and Figures

\begin{figure}[p]
\fbox{\includegraphics[width=15cm]{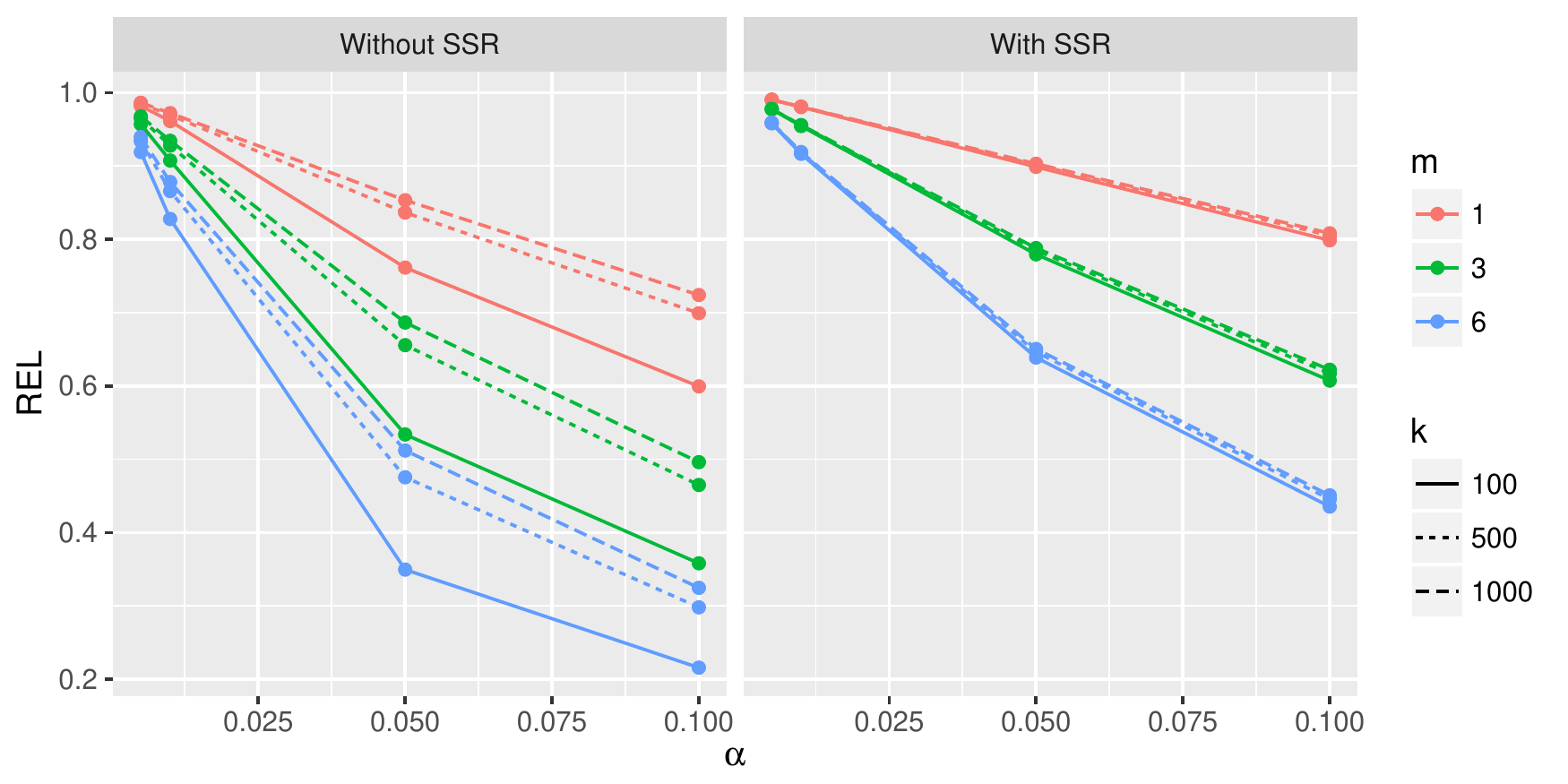}}
\caption{Values of $REL$ for varying values of $k$, $m$ and $\alpha$.  Left-panel shows results with no power requirement (i.e., $A = (1-psp)^{m}$).  Right-panel shows results with power requirement (i.e., $A$ defined as per equation \ref{Apwr} (with $c_{50}=0.5$ and $c_{95}=0.8$)).}  
\label{fig:REL}
\end{figure}

\begin{figure}[p]
\fbox{ \includegraphics[width=15cm]{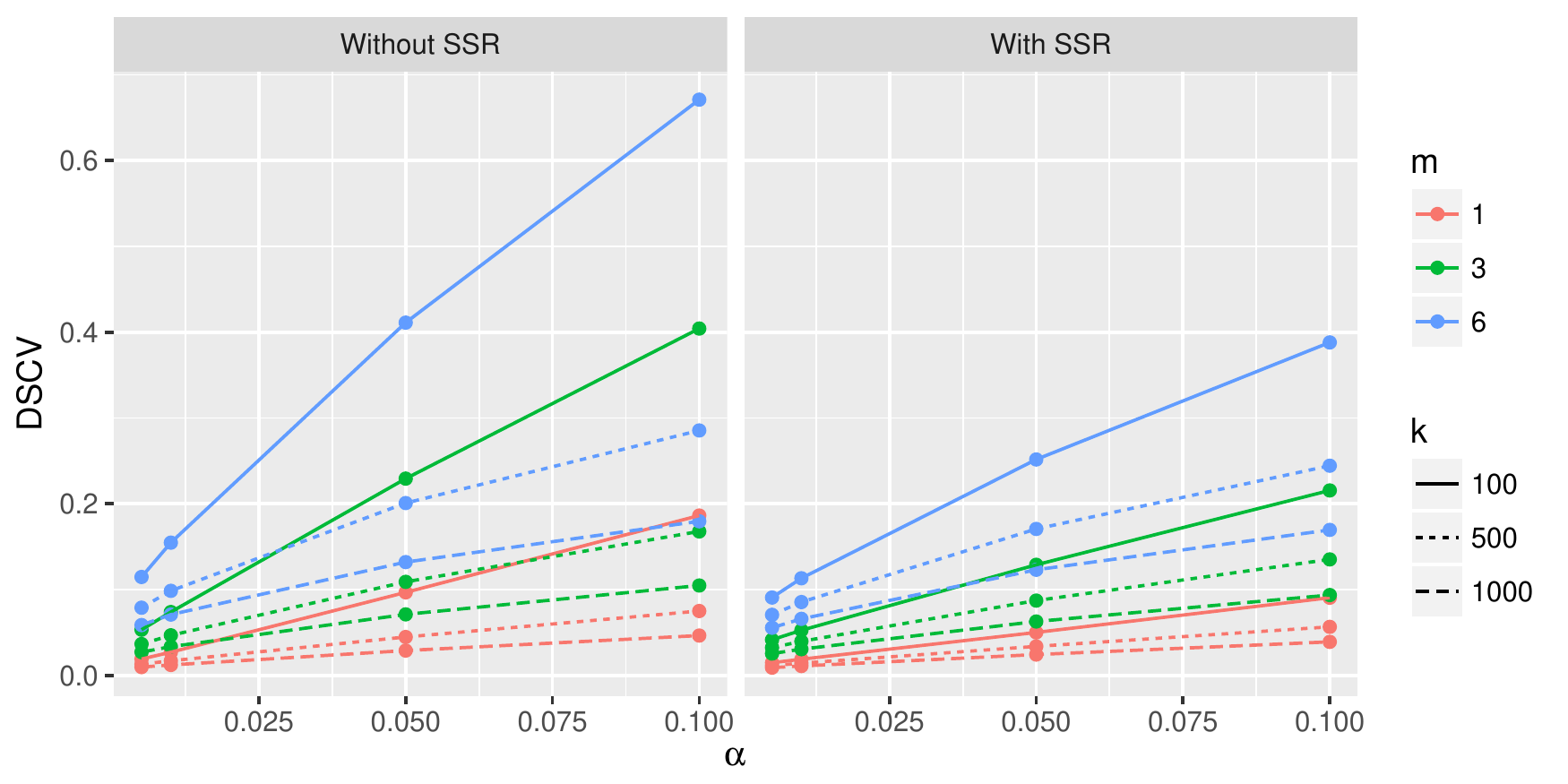}}
\caption{Values of $DSCV$ for varying values of $k$, $m$ and $\alpha$.  Left-panel shows results with no power requirement (i.e., $A = (1-psp)^{m}$).  Right-panel shows results with power requirement (i.e., $A$ defined as per equation \ref{Apwr} (with $c_{50}=0.5$ and $c_{95}=0.8$)).} 
\label{fig:DSCV}
\end{figure}

\begin{figure}[p]
\fbox{\includegraphics[width=15cm]{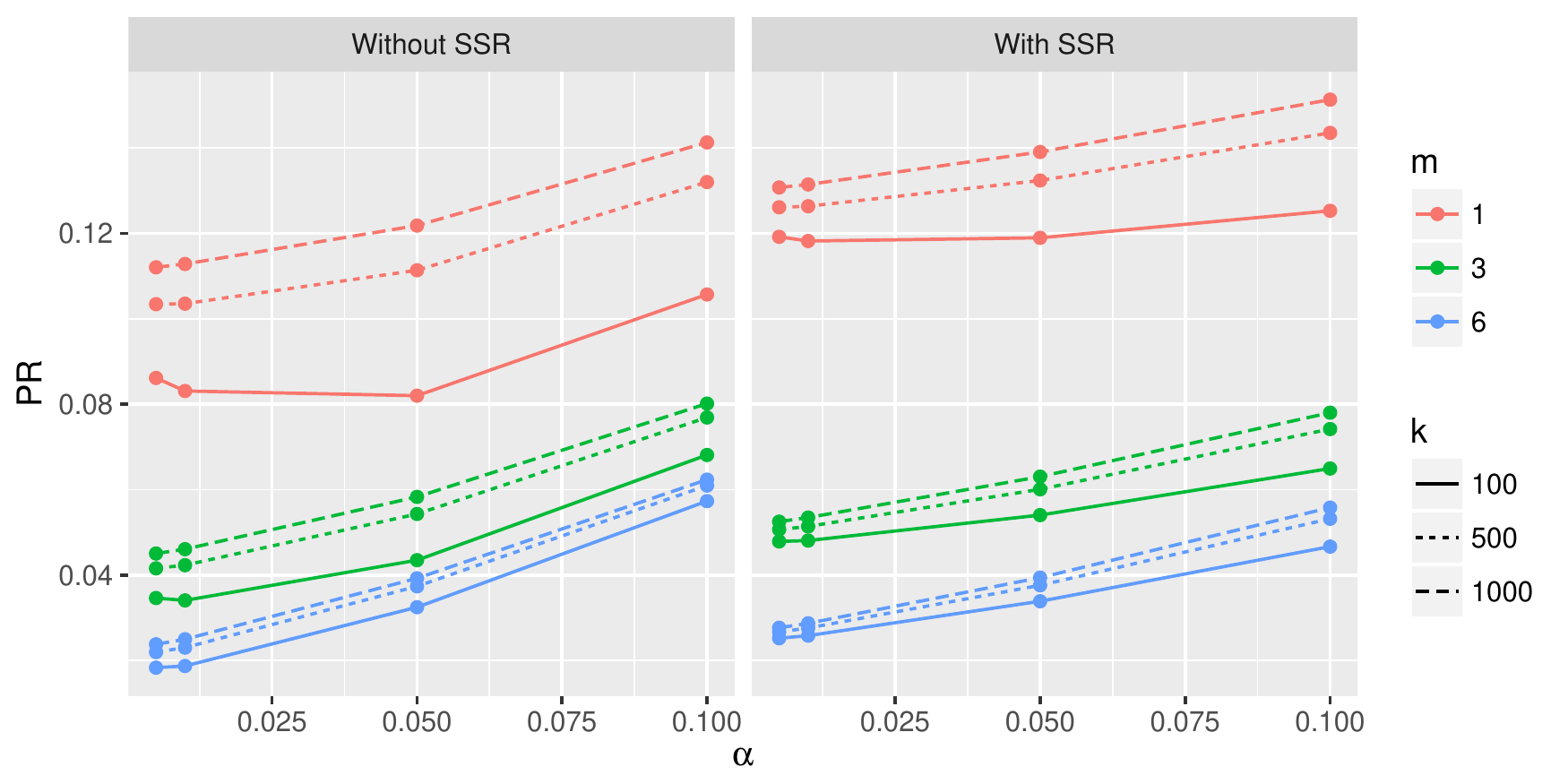}}
\caption{Values of the publication rate (PR), equal to $N_{PUB}/N_{ATM}$, for varying values of $k$, $m$ and $\alpha$.  Left-panel shows results with no power requirement (i.e., $A = (1-psp)^{m}$).   Right-panel shows results with power requirement (i.e., $A$ defined as per equation \ref{Apwr} (with $c_{50}=0.5$ and $c_{95}=0.8$)).} 
\label{fig:PR}
\end{figure}

\begin{figure}[p]
\fbox{\includegraphics[width=15cm]{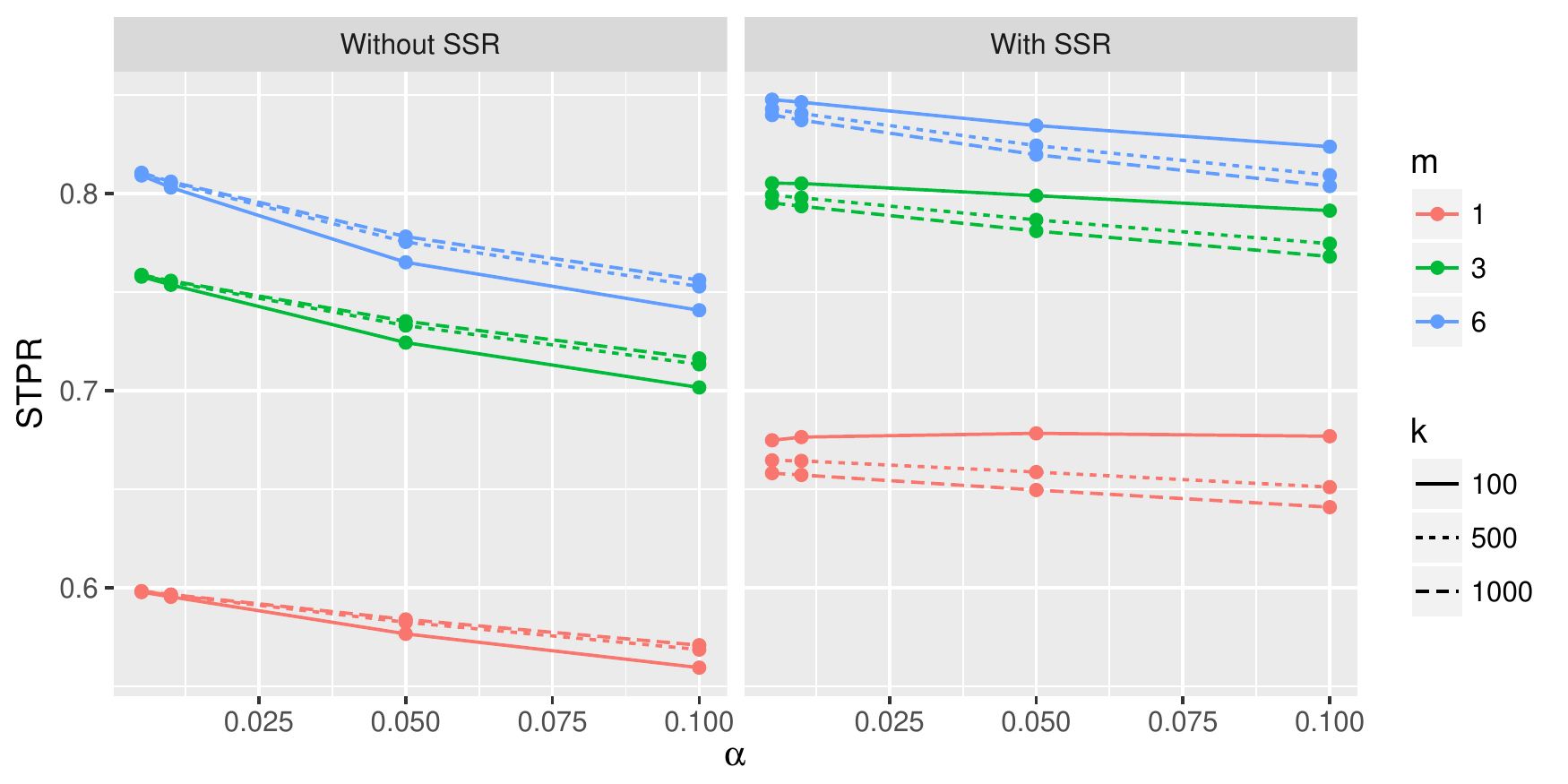}}
\caption{Values of $DSCV$ for varying values of $k$, $m$ and $\alpha$.  Left-panel shows results with no power requirement (i.e., $A = (1-psp)^{m}$).   Right-panel shows results with power requirement (i.e., $A$ defined as per equation \ref{Apwr} (with $c_{50}=0.5$ and $c_{95}=0.8$)).} 
\label{fig:silenced}
\end{figure}

\begin{figure}[p]
\fbox{\includegraphics[width=15cm]{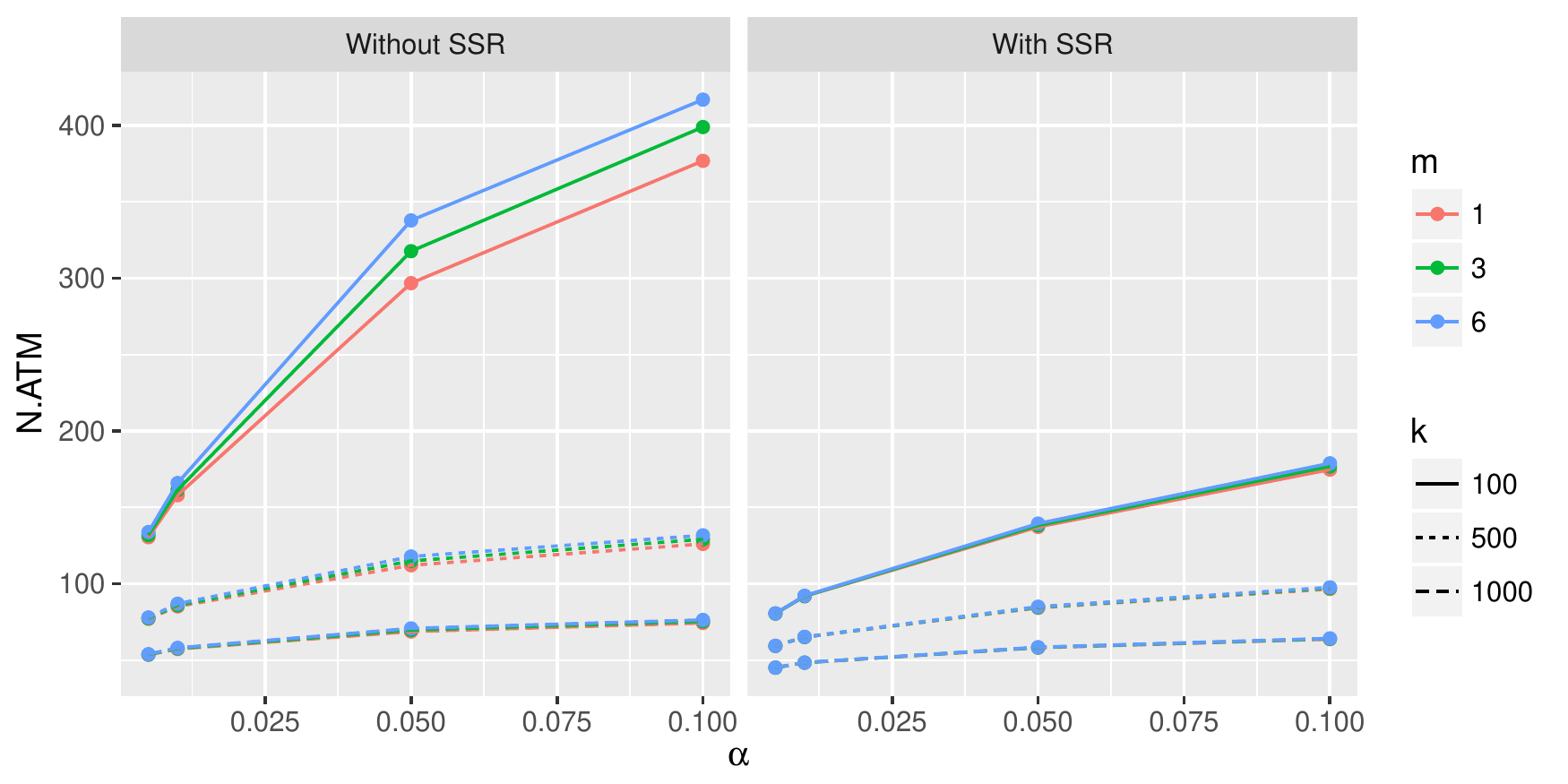}}
\caption{Values of $N_{ATM}$ for varying values of $k$, $m$ and $\alpha$.  Left-panel shows results with no power requirement (i.e., $A = (1-psp)^{m}$).   Right-panel shows results with power requirement (i.e., $A$ defined as per equation \ref{Apwr} (with $c_{50}=0.5$ and $c_{95}=0.8$)).} 
\label{fig:Natm}
\end{figure}

\begin{figure}[p]
\fbox{\includegraphics[width=15cm]{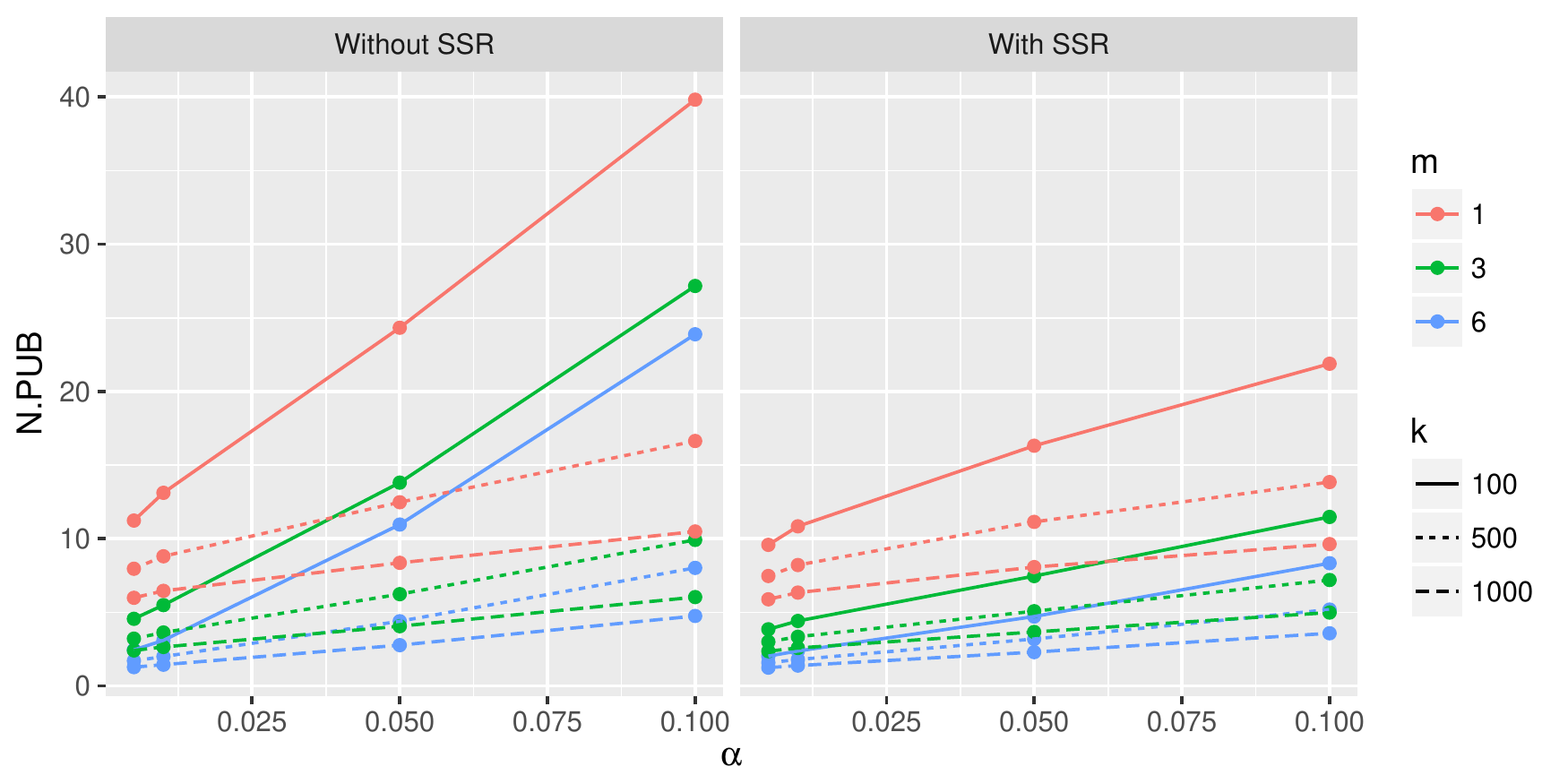}}
\caption{Values of $N_{PUB}$ for varying values of $k$, $m$ and $\alpha$.  Left-panel shows results with no power requirement (i.e., $A = (1-psp)^{m}$).   Right-panel shows results with power requirement (i.e., $A$ defined as per equation \ref{Apwr} (with $c_{50}=0.5$ and $c_{95}=0.8$)).} 
\label{fig:Npub}
\end{figure}

\begin{figure}[p]
\fbox{\includegraphics[width=15cm]{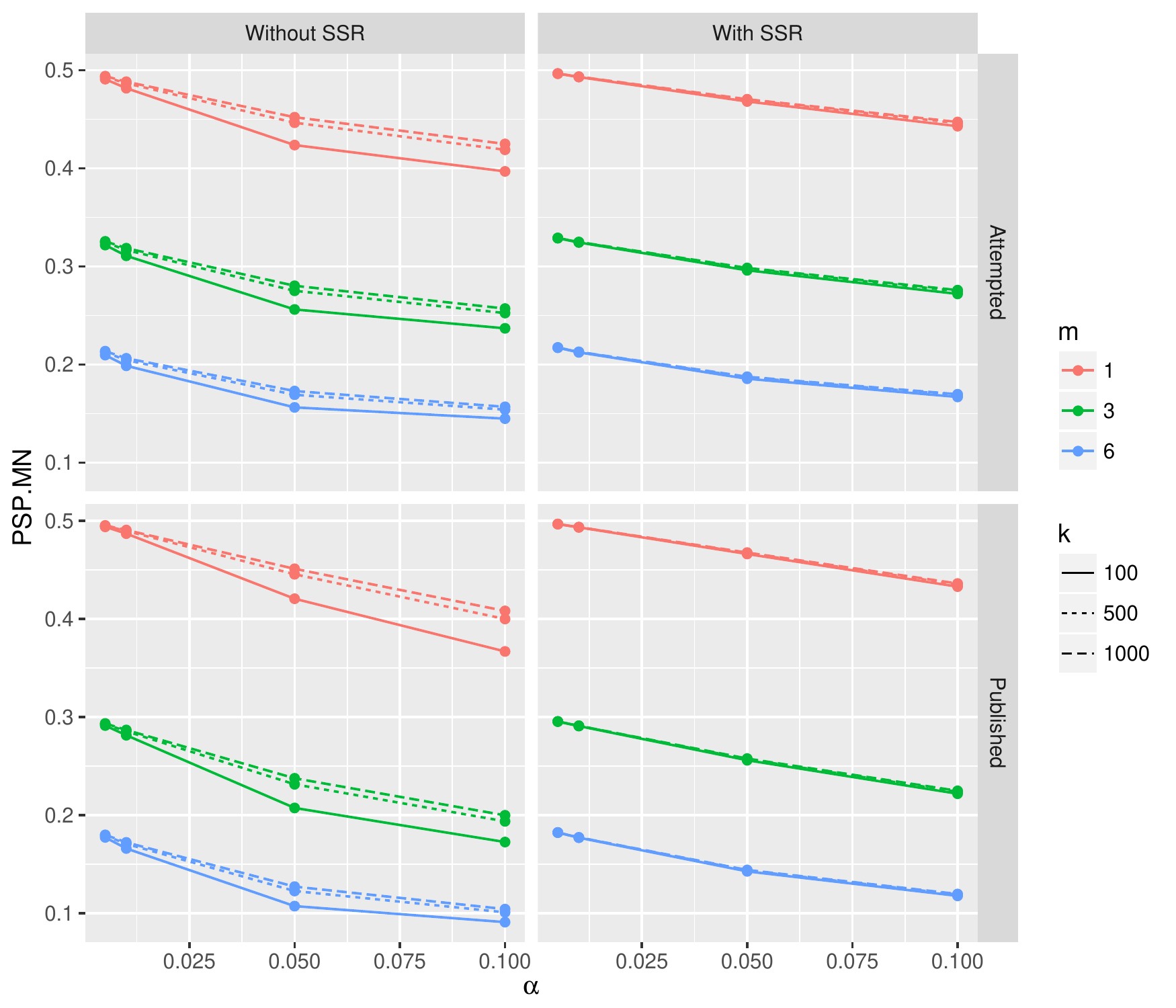}}
\caption{Values of mean $PSP$ for both attempted and published studies.  Ecosystems have varying values of $k$, $m$ and $\alpha$.  Left-panel shows results with no power requirement (i.e., $A = (1-psp)^{m}$).   Right-panel shows results with power requirement (i.e., $A$ defined as per equation \ref{Apwr} (with $c_{50}=0.5$ and $c_{95}=0.8$)).} 
\label{fig:meanpsp}
\end{figure}

\begin{figure}[p]
\fbox{\includegraphics[width=15cm]{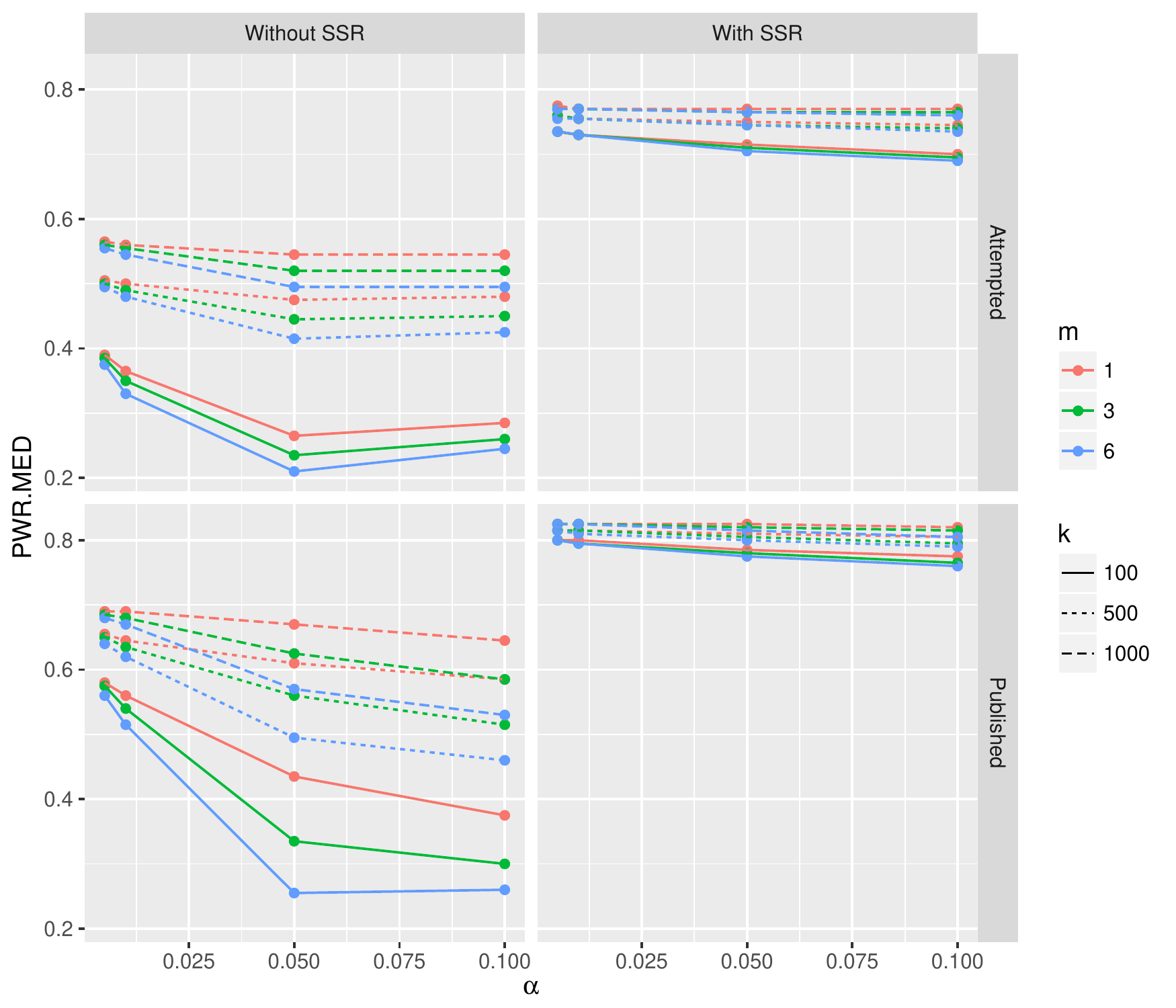}}
\caption{Values of median $PWR$ for both attempted and published studies.  Ecosystems have varying values of $k$, $m$ and $\alpha$.  Left-panel shows results with no power requirement (i.e., $A = (1-psp)^{m}$).   Right-panel shows results with power requirement (i.e., $A$ defined as per equation \ref{Apwr} (with $c_{50}=0.5$ and $c_{95}=0.8$)).} 
\label{fig:medpwr}
\end{figure}

\begin{figure}[p]
\fbox{\includegraphics[width=15cm]{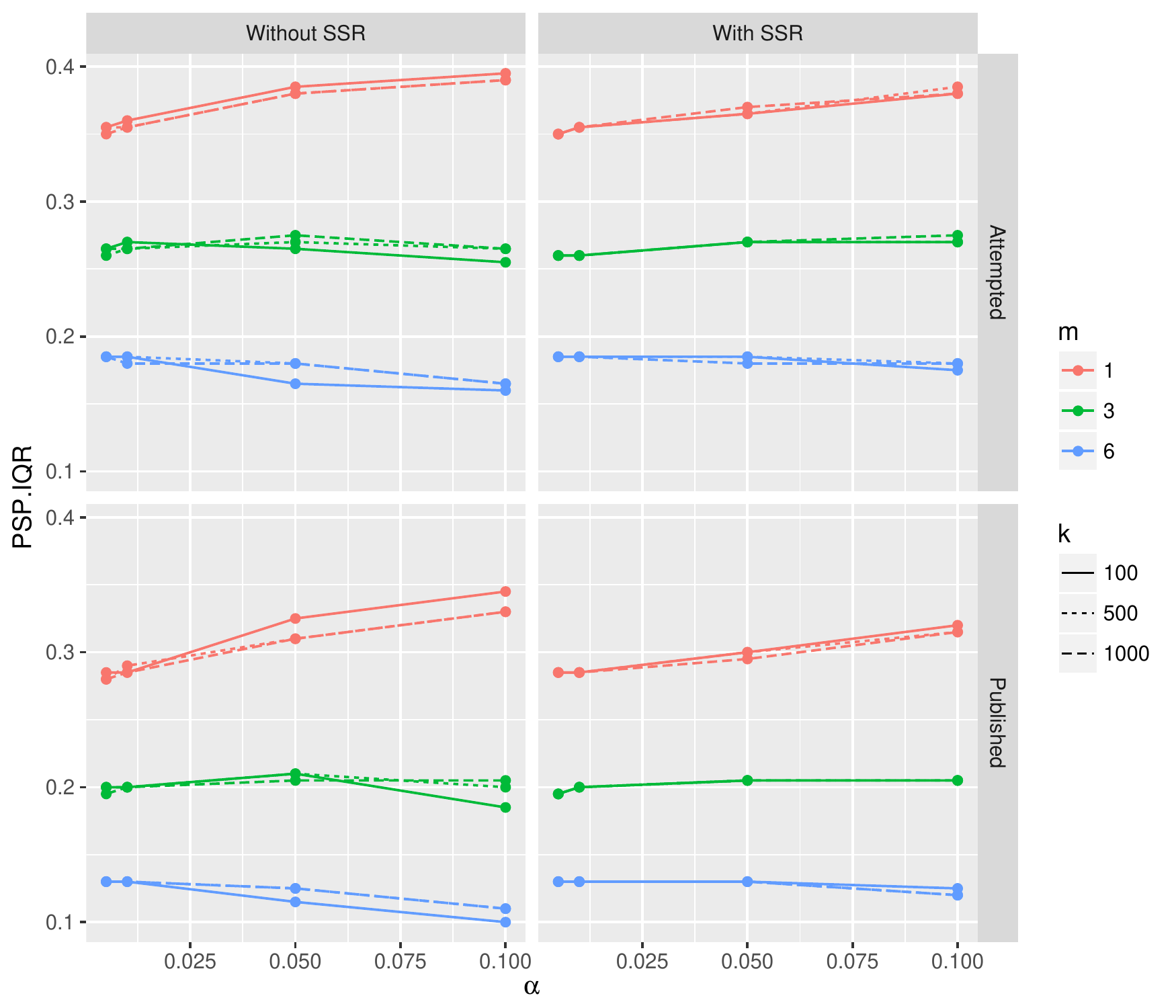}}
\caption{The inter-quartile range of $PSP$ values for both attempted and published studies.  Ecosystems have varying values of $k$, $m$ and $\alpha$.  Left-panel shows results with no power requirement (i.e., $A = (1-psp)^{m}$).   Right-panel shows results with power requirement (i.e., $A$ defined as per equation \ref{Apwr} (with $c_{50}=0.5$ and $c_{95}=0.8$)).} 
\label{fig:IQpsp}
\end{figure}

\begin{table}[ht]
\centering
\begin{tabular}{rrrrrrr}
  \hline
$\alpha$ & $k$ & $m$ & $PSP_{ATM}$ & $PSP_{PUB}$ & $PWR_{ATM}$ & $PWR_{PUB}$ \\ 
  \hline
0.005 & 100 & 1 & [0.32 , 0.67] & [0.35 , 0.64] & [0.21 , 0.62] & [0.39 , 0.76] \\ 
  0.005 & 100 & 3 & [0.18 , 0.44] & [0.18 , 0.38] & [0.20 , 0.61] & [0.38 , 0.76] \\ 
  0.005 & 100 & 6 & [0.10 , 0.29] & [0.10 , 0.24] & [0.20 , 0.61] & [0.36 , 0.75] \\ 
  0.005 & 500 & 1 & [0.32 , 0.67] & [0.36 , 0.64] & [0.31 , 0.72] & [0.47 , 0.81] \\ 
  0.005 & 500 & 3 & [0.18 , 0.44] & [0.18 , 0.38] & [0.30 , 0.71] & [0.46 , 0.81] \\ 
  0.005 & 500 & 6 & [0.11 , 0.29] & [0.10 , 0.24] & [0.29 , 0.71] & [0.45 , 0.80] \\ 
  0.005 & 1000 & 1 & [0.32 , 0.67] & [0.36 , 0.64] & [0.36 , 0.76] & [0.51 , 0.84] \\ 
  0.005 & 1000 & 3 & [0.18 , 0.45] & [0.19 , 0.38] & [0.35 , 0.76] & [0.50 , 0.84] \\ 
  0.005 & 1000 & 6 & [0.11 , 0.29] & [0.11 , 0.24] & [0.34 , 0.75] & [0.49 , 0.84] \\ 
  0.050 & 100 & 1 & [0.22 , 0.61] & [0.26 , 0.58] & [0.14 , 0.48] & [0.23 , 0.66] \\ 
  0.050 & 100 & 3 & [0.11 , 0.38] & [0.09 , 0.30] & [0.12 , 0.44] & [0.16 , 0.58] \\ 
  0.050 & 100 & 6 & [0.06 , 0.22] & [0.04 , 0.15] & [0.11 , 0.40] & [0.12 , 0.49] \\ 
  0.050 & 500 & 1 & [0.26 , 0.64] & [0.29 , 0.60] & [0.28 , 0.68] & [0.41 , 0.78] \\ 
  0.050 & 500 & 3 & [0.12 , 0.40] & [0.12 , 0.32] & [0.25 , 0.66] & [0.34 , 0.75] \\ 
  0.050 & 500 & 6 & [0.06 , 0.24] & [0.05 , 0.18] & [0.22 , 0.64] & [0.28 , 0.71] \\ 
  0.050 & 1000 & 1 & [0.26 , 0.64] & [0.29 , 0.60] & [0.34 , 0.75] & [0.47 , 0.83] \\ 
  0.050 & 1000 & 3 & [0.13 , 0.40] & [0.12 , 0.33] & [0.31 , 0.73] & [0.41 , 0.80] \\ 
  0.050 & 1000 & 6 & [0.07 , 0.25] & [0.06 , 0.18] & [0.28 , 0.71] & [0.35 , 0.77] \\ 
   \hline
\end{tabular}
\caption{The inter-quartile ranges ([25th,75th]) of PSP and PWR values for attempted (ATM) and published (PUB) studies without SSR. Ecosystems are defined by $\alpha=0.05$ or $\alpha=0.005$, and $A = (1-psp)^{m}$, $B=0$, with varying values of $k$ and $m$ as indicated.}
\label{tab:IQ005to05}
\end{table}

\begin{table}[ht]
\centering
\begin{tabular}{rrrrrrr}
  \hline
$\alpha$ & $k$ & $m$ & $PSP_{ATM}$ & $PSP_{PUB}$ & $PWR_{ATM}$ & $PWR_{PUB}$ \\ 
  \hline
  0.005 & 100 & 1 & [0.32 , 0.67] & [0.36 , 0.64] & [0.64 , 0.84] & [0.72 , 0.89] \\ 
  0.005 & 100 & 3 & [0.19 , 0.45] & [0.19 , 0.38] & [0.64 , 0.84] & [0.72 , 0.89] \\ 
  0.005 & 100 & 6 & [0.12 , 0.30] & [0.11 , 0.24] & [0.63 , 0.84] & [0.72 , 0.89] \\ 
  0.005 & 500 & 1 & [0.32 , 0.67] & [0.36 , 0.64] & [0.66 , 0.86] & [0.73 , 0.90] \\ 
  0.005 & 500 & 3 & [0.19 , 0.45] & [0.19 , 0.38] & [0.66 , 0.86] & [0.73 , 0.90] \\ 
  0.005 & 500 & 6 & [0.12 , 0.30] & [0.11 , 0.24] & [0.65 , 0.86] & [0.73 , 0.90] \\ 
  0.005 & 1000 & 1 & [0.32 , 0.67] & [0.36 , 0.64] & [0.66 , 0.88] & [0.74 , 0.90] \\ 
  0.005 & 1000 & 3 & [0.19 , 0.45] & [0.19 , 0.38] & [0.66 , 0.87] & [0.74 , 0.90] \\ 
  0.005 & 1000 & 6 & [0.12 , 0.30] & [0.11 , 0.24] & [0.66 , 0.87] & [0.74 , 0.90] \\ 
  0.050 & 100 & 1 & [0.28 , 0.65] & [0.32 , 0.61] & [0.61 , 0.82] & [0.70 , 0.87] \\ 
  0.050 & 100 & 3 & [0.15 , 0.42] & [0.14 , 0.35] & [0.61 , 0.82] & [0.70 , 0.87] \\ 
  0.050 & 100 & 6 & [0.08 , 0.26] & [0.07 , 0.20] & [0.60 , 0.82] & [0.69 , 0.86] \\ 
  0.050 & 500 & 1 & [0.28 , 0.65] & [0.32 , 0.61] & [0.64 , 0.85] & [0.72 , 0.89] \\ 
  0.050 & 500 & 3 & [0.15 , 0.42] & [0.14 , 0.35] & [0.64 , 0.85] & [0.72 , 0.89] \\ 
  0.050 & 500 & 6 & [0.08 , 0.26] & [0.07 , 0.20] & [0.64 , 0.85] & [0.71 , 0.89] \\ 
  0.050 & 1000 & 1 & [0.28 , 0.66] & [0.32 , 0.61] & [0.66 , 0.87] & [0.74 , 0.90] \\ 
  0.050 & 1000 & 3 & [0.15 , 0.42] & [0.14 , 0.35] & [0.66 , 0.87] & [0.73 , 0.90] \\ 
  0.050 & 1000 & 6 & [0.08 , 0.26] & [0.07 , 0.20] & [0.66 , 0.87] & [0.73 , 0.90] \\ 
   \hline
\end{tabular}
\caption{The inter-quartile ranges ([25th,75th]) of PSP and PWR values for attempted (ATM) and published (PUB) studies with SSR. Ecosystems are defined by $\alpha=0.05$ or $\alpha=0.005$, and $A$ defined as per equation \ref{Apwr} (with $c_{50}=0.5$ and $c_{95}=0.8$), with varying values of $k$ and $m$ as indicated.}
\label{tab:IQ005to05}
\end{table}

\end{description}

\end{document}